\documentclass[final,numberedheadings] {aipproc} 
\usepackage{psfrag}
\usepackage{amsmath,amssymb}
\usepackage{color}

\layoutstyle{6x9}
\graphicspath{{eps/}}
\newcommand{\rme}{\mathrm{e}}
\newcommand{\rmc}{\mathrm{c}}
\newcommand{\rmd}{\mathrm{d}}
\newcommand{\rmH}{\mathrm{H}}
\newcommand{\rmi}{\mathrm{i}}
\newcommand{\rmq}{\mathrm{q}}
\newcommand{\rms}{\mathrm{s}}

\newcommand{\rmIm}{\mathop{\mathrm{Im}}\nolimits}
\newcommand{\rmRe}{\mathop{\mathrm{Re}}\nolimits}
\newcommand{\eref}[1]{eq.~(\ref{#1})} 
\newcommand{\fref}[1]{fig.~\ref{#1}}  
\newcommand{\Eref}[1]{Eq.~(\ref{#1})} 
\newcommand{\Fref}[1]{Fig.~\ref{#1}}  
\newcommand{\Or}{\mathord{\mathrm{O}}}
\providecommand{\openone}{\leavevmode\hbox{\small1\kern-3.8pt\normalsize1}}

\newcommand{\tr}{\mathop{\mathrm{tr}}\nolimits}

\newcommand{\mcU}{\mathcal{U}}
\newcommand{\mcH}{\mathcal{H}}

\def\>{\rangle}
\def\<{\langle}

\newcommand{\Real}{\mathbb{R}}

\newcommand{\ie}{\textit{i.e.} }

\newcommand{\into}{\int_0^t \rmd \tau \int_0^t \rmd \tau'}
\def\braket#1#2{\< #1 | #2 \>}

\newcounter{hunger}
\newcommand{\e}{\rme}
\newcommand{\la}{\langle}
\newcommand{\ra}{\rangle}

\newcommand{\eps}{\epsilon}

\begin{document}

\title[RMT:  Decoherence, and fidelity decay]{Random matrix models for
decoherence and fidelity decay in quantum information systems}


\author{Carlos Pineda}{ }
\author{Thomas H. Seligman}
{address={Instituto de Ciencias F\'isicas, Universidad Nacional Aut\'onoma
de M\'exico, Cuernavaca, M\'exico\\ Centro Internacional de Ciencias, Cuernavaca, M\'exico}}

\begin{abstract}
This course aims to introduce the student to random matrix models for
decoherence and fidelity decay. They present a very powerful alternate
approach, that emphasizes the disordered character of many environments and
uncontrollable perturbations/couplings.  The inherent integrability of such
models makes analytic studies possible. We limit our considerations to linear
response treatment, as high fidelity and small decoherence are the backbone of
quantum information processes.  For fidelity decay, where experimental results
are available, a comparison with experiments shows excellent agreement with
random matrix theory predictions.
\end{abstract}

\maketitle


\section{Introduction}

In this course we shall discuss how random matrix theory (RMT) can be used to
describe the two main sources of errors in a quantum information process:
\begin{list}{(\Alph{hunger})}{\usecounter{hunger}}
\item
Loss of exactitude due to inherent errors of the physical reproduction of the
algorithm. 
\item
Loss of coherence due to coupling to and entanglement with some outer system.
\end{list}
We thus distinguish errors in the unitary time evolution from those caused by
the loss of unitarity due to external action caused by the environment,
from which perfect isolation is never possible. As we shall see in the third
section such considerations are intimately related to the possibility of
reverting some time evolution by a time reflection of the state, \ie to the
old problem of de-equilibration after time reversal, usually associated to
the name of Loschmidt in his discussions with Boltzmann, about 130 years ago
\cite{Loschmidt:76}. Yet Lord Kelvin \cite{Thompson:74}, some years earlier,
gave an excellent account of the problem:

\begin{quote}
\it ``If
we allowed this equalization to proceed for a certain time, and then reversed
the motions of all the molecules, we would observe a disequalization.
However, if the number of molecules is very large, as it is in a gas, any
slight deviation from absolute precision in the reversal will greatly shorten
the time during which disequalization occurs\ldots Furthermore, if we take
account of the fact that no physical system can be completely isolated from
its surroundings but is in principle interacting with all other molecules in
the universe, and if we believe that the number of these latter molecules is
infinite, then we may conclude that it is impossible for
temperature-differences to arise spontaneously\ldots''
\end{quote}
Though speaking in a classical context, he clearly points to the two problems
affecting the stability of a quantum information process: precision and
uncontrollable interaction with the environment. What is more, we are warned
that the separation is artificial, as the second implies the first. He
further points out the essentially statistical character of perturbations and
coupling. This indicates why we should use a randomized description.
Decoherence in terms of spontaneous emission, is the intellectual basis for
any model based on harmonic oscillators as environment.  This assumes a very
unusual isolation of the system, maybe realistic for the decay of excitations
of atoms and molecules in interstellar space.

The use of random matrix theory (RMT) to understand quantum systems started
modestly in the restricted field of low - energy nuclear physics, though it
was introduced by no one less than  Eugene Wigner \cite{Wig51}. Since then,
it has evolved to applications reaching from mesoscopics, molecules, atoms
all the way to elementary particles \cite{guhr98random}. Early proposals of
applications to correlation analysis  of times series \cite{wishartRMT} have
since developed as an important tool from biology \cite{RMTbiology} to
econo-physics \cite{econophysicsPotters, econophysicsLillo}, and the
techniques have merged with those developed for quantum systems
\cite{guhr98random}.

RMT models for decoherence have been introduced some time ago \cite{caoslutz,
GPS04, Gorin2003} and more recently such a model was proposed for
fidelity decay \cite{1367-2630-6-1-020} and successfully compared to
experiment \cite{1367-2630-7-1-152, expRudi, gorin-weaver}.  More recently
the models and calculation techniques have been improved \cite{reflosch} and
specialized for the treatment of qubits and quantum registers
\cite{pinedaRMTshort, pinedalong}.

The purpose of this course is to give an introduction to the application of
RMT both to fidelity decay and to decoherence. We shall start in the next
section with a short overview of RMT. In section three we shall give an
introduction to fidelity decay and present the RMT model we intend to use.
We then discuss linear response in this context and proceed to calculate
fidelity for the RMT model in this approximation. A comparison with exact
solutions and experiments follows and we close with a special case of reduced
fidelity decay known as the fidelity freeze.  In section four we shall
concentrate more on quantum information systems, choosing as a central system
a quantum register of $n$ qubits. First we shall discuss some general results
that can be derived using the linear response approximation, and then perform
explicit calculations for a random matrix model. To end this section, we give
heuristic results for the relation between loss of coherence and loss of
internal entanglement in the central system.  In the concluding remarks we
dwell on the fact, that chaos can slow down fidelity decay and decoherence.

\section{Random Matrix Theory: an overview}

We shall give here a very concise overview of the concepts and definitions we
use, and some properties of the classical random matrix ensembles as
introduced by Cartan \cite{cartanRMT}.   For a wider description we refer the
reader to the standard textbook by Mehta \cite{mehta}. A modern review is
given in \cite{guhr98random}.

To construct an {\it ensemble} of random Hamiltonian matrices of minimal
information we start out with the {\it set} of Hermitian matrices and require
that the ensemble be independent of the basis chosen in the $N$-dimensional
Hilbert space. This implies, that this measure must be invariant under
similarity transformations by the unitary group $\mcU (N)$. The measure
$\rmd\mu$ thus fulfills
\begin{equation}
\rmd\mu (H) = \rmd\mu (UHU^\dagger), \quad U \in \mcU (N).
\end{equation}
Checking numbers of conditions, to fix the measure, against the number of
variables, we find that there are still $N$ missing conditions. These
correspond to the eigenvalues of the hermitian matrix $H$, and up to this
point are variables, whose distribution is not determined.  
Balian shows \cite{balian} that the minimum
information requirement is sufficient to fix this. His argument leads to
matrix elements $H_{m,n}$ being Gaussian distributed both in their real and
imaginary part, all with the same width, and independent except for the
Hermiticity condition. \ie,
\begin{equation}
\rmd\mu(H) \propto \rme^{-\alpha^2 \tr(H^2)/2} \rmd \rmRe (H) \, \rmd \rmIm (H).
\end{equation}
This result can be rewritten in terms of the Haar measure of the symmetry
group [in this case $\mcU (N)$] and a measure for the eigenenergies $E_i$ of
$H$.  This idea can be generalized to groups other than $\mcU (N)$.  We
thus allow arbitrary invariance groups ${\mathcal G}_\beta$.  The measure
then reads as
\begin{equation}\label{eq:measuredecomposition}
\rmd\mu_\beta (H) \propto \rmd h_\beta (G)  \prod_{i \ne j}
       \exp \vert E_i - E_j \vert ^\beta \, \prod_l \rmd E_l,
        \quad G\in {\mathcal G}_\beta
\end{equation}
where $\rmd h_\beta(G)$ indicates the invariant Haar measure of the group
${\mathcal G}_\beta$.

The alternate groups are the orthogonal group ${\cal O}(N)$, labeled with
$\beta = 1$, and the unitary symplectic group $\mathcal{U \!\!\! S \!\!\!
P}(2N)$, labeled with $\beta=4$, in which case we get the Gaussian orthogonal
and the Gaussian symplectic ensembles (GOE, GSE). The former is an ensemble
of real symmetric Hamiltonians appropriate for describing most time reversal
invariant systems, except such of semi-integer spin where we have to use the
latter.  If we use the unitary group ($\beta=2$) we obtain the Gaussian
unitary ensemble (GUE), appropriate for describing time-reversal breaking
systems.  Note that the spectral measure is analytic for the GUE and GSE
($\beta=2$ and $4$) but not for the GUE ($\beta=1$); this leads to the fact
that analytic calculations, in the GOE, are often more complicated than for
the other two ensembles.

It is possible to construct the equivalent ensembles of unitary matrices
giving rise to the circular orthogonal, unitary and symplectic ensembles
(COE, CUE and CSE). These are unitary matrices with the same invariance
properties under similarity transformations. Yet they have larger symmetries
which define the ensembles completely. Thus the CUE is actually invariant
under $\mcU(N) \times \mcU(N)$ as
\begin{equation}
\rmd\nu (S) = \rmd\nu (USV)
\end{equation}
where $U$ and $V$ run over the group independently.  This defines the
invariant Haar measure of the unitary group, \ie, the CUE is the unitary
group plus its Haar measure.  The COE on the other hand consists of unitary
symmetric matrices that do NOT form a group, and its invariance group is the
unitary group, but not as a similarity transformation. Rather we have
\begin{equation}
\rmd\nu (S) = \rmd\nu (USU^t),
\end{equation}
where $ ^t $ indicates transposition. While the transposed is not the inverse
of a unitary transformation, this is the case for the orthogonal subgroup
thus displaying the same symmetry under similarity transformations as the
GOE.  The circular ensembles (CUE, COE, and CSE) are appropriate to describe
 unitary evolution operators such as the
scattering matrix or the Floquet operator corresponding to a periodic time
dependence of a Hamiltonian.

Note that for any of the three invariance groups we can construct generalized
invariant ensembles with arbitrary spectral distributions inserted in
\eref{eq:measuredecomposition}. Examples include the Poisson unitary and
orthogonal ensembles (PUE, POE) \cite{Dittes}, where a random uniform
distribution (in an appropriate interval) of eigenvalues is assumed. This
ensemble mainly serves as contrast to display the effect of spectral
correlations. An equidistant ``picket fence'' spectrum can also be inserted to
take spectral stiffness to the extreme.

It thus becomes clear, that the characteristic properties of the classical
ensembles are found in their spectral statistics, and this has been used
extensively. The level density is of limited interest, as this is really a
specific property of each system. The classical ensembles of Hamiltonians
lead in, the large $N$ limit, to a semicircular level density, which is quite
unrealistic. The ensembles of unitary matrices on the other hand lead to
constant densities of eigenphases on the unit circle, which is often
realistic.  If analytic solutions are thought they refer mostly to the center
of the spectrum, where the level density is flat and only needs to be
normalized to one to study spectral correllations. If spectral
statistics of some experiment are compared with RMT, the spectrum must be
unfolded, \ie the spectrum must be normalized locally. For numerical
simulations one can take advantage of the generalized invariant ensembles and
introduce unfolded spectra in the calculations.

\begin{figure}\label{fig:psandktwo}
\psfrag{b1}{\footnotesize $\beta_1$, GUE}
\psfrag{b2}{\footnotesize $\beta_2$, GOE}
\psfrag{b4}{\footnotesize $\beta_4$, GSE}
\psfrag{Ps}{$P(s)$}
\psfrag{t}{$t/\tau_\rmH$}
\psfrag{Ktwo}{$1-b_2(t)$}
\psfrag{s}{$s$}
\includegraphics{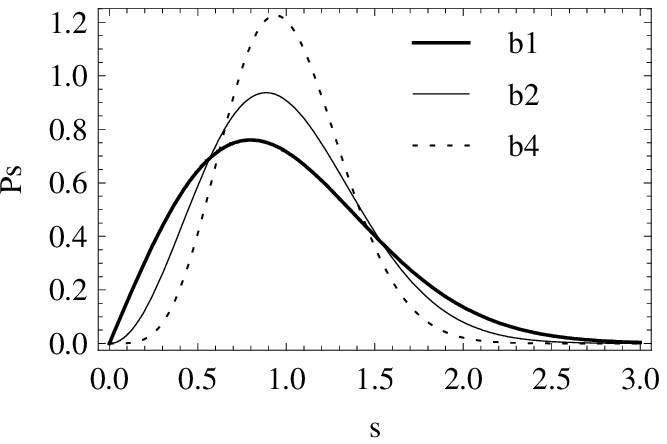}\hspace{1cm}\includegraphics{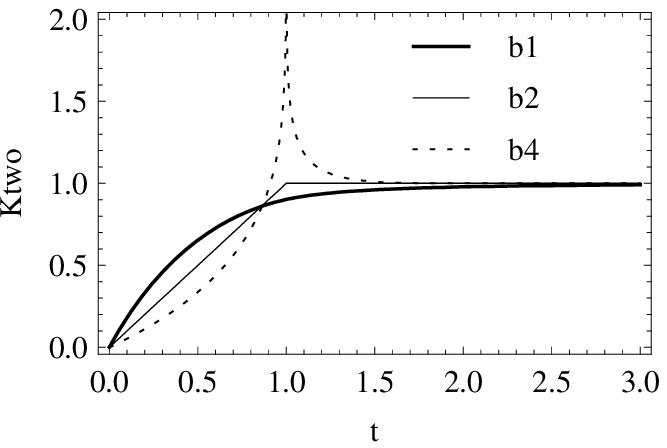}
\caption{The Wigner surmise for the nearest neighbor spacing distribution
$P(s)$ and the spectral form factor are shown on the left and right panel
  respectively for the three ensembles discussed.}
\end{figure}

The most popular statistical test is the nearest neighbor spacing
distribution \ie the distribution of spacings of neighboring levels in the
unfolded spectrum.  This function depends in principle on all $n$-point
functions, but its short range behavior is dominated by two-point function.
The Wigner surmise gives the simple estimate
\begin{equation}
P_\beta (s) \propto s^\beta \rme^{-\gamma_\beta s^2},\quad
\gamma_\beta=\frac{\Gamma^2[(\beta+2)/2]}{\Gamma^2[(\beta+1)/2]}
\end{equation}
for the nearest neighbor  spacing distribution, which is exact for  $n=2$
\cite{guhr98random}.  Fig.~\ref{fig:psandktwo}(a) shows this surmise which is
quite good even for the large $N$.

Another quantity of considerable interest and that will be used often
during this course is the form factor
\begin{equation}
 K_2(t) = \frac{1}{N} \sum_{i,j} \rme^{\rmi t (E_i-E_j)}
 = \frac{1}{N} \left| \sum_{i} \rme^{\rmi t E_i} \right|^2,
\label{eq:formfactordef}
\end{equation}
\ie of the two point function in the time domain.  Here $E_i$ are the
eigenenergies of the Hamiltonian under consideration.  We define the spectral
form factor for an ensemble in terms of the ensemble average $\langle K_2
\rangle $ as $b_2(t/\tau_\rmH) =  1+ \delta(t/\tau_\rmH)-\langle
K_2(t/\tau_\rmH)\rangle$ in terms of a dimensionless time $\tau=t/\tau_\rmH$
whose scale is set by the Heisenberg time $\tau_\rmH=2 \pi \hbar/\text{(mean
level spacing)}$.  Exact and approximate analytic results for $b_2$ are known
in the large $N$ limit. For the GUE the exact form is quite simple and given by
\begin{equation}
b_2^{(2)}(\tau)=\begin{cases} 1-|\tau|& \text{if\ }|\tau|\le 1,\\
0& \text{if\ }|\tau|>1 \end{cases}.
\label{eq:b2defGUE}
\end{equation}
It is of interest to remark, that for the GSE case, $b_2$ has a singularity
at Heisenberg time, while for GUE and GOE respectively the first and third
derivatives respectively are discontinuous.  
Note finally that, often, $K_2$ is called the
spectral form factor.  We reserves this term for $b_2$.

\section{Random matrix theory of fidelity decay}
\label{R} As, in this school, the stability of unitary evolution has received
less attention, we have to introduce the problem in more detail, define
fidelity and mention why it is a sensible benchmark.  We assume an ideal
evolution $U$ and a physical evolution $U_\eps$.  To estimate the deviation
between these two evolutions at the end of, or at any intermediate time
(step) of the quantum evolution, a convenient and widely used measure of this
deviation is the overlap of the relevant register function under the two time
evolutions, \ie their cross correlation
\begin{equation}\label{eq:definefidelityecho}
f(t) = \<\psi \vert M_\eps(t) \vert \psi\>
\end{equation}
also known as the fidelity amplitude. Here
\begin{equation}\label{eq:defEcho}
 M_\eps(t)= U_0^\dagger (t) \,U_\eps (t) = U_0 (-t) \,U_\eps (t)
\end{equation}
is known as the echo operator, so $f(t)=[\<\psi|U_0^\dagger(t)]
[U_\epsilon(t)|\psi\>]$. This name results from the fact that we can
reinterpret the definition of the fidelity amplitude as an auto-correlation
function of evolution under the echo-operator, \ie forward evolution with one
Hamiltonian and backward (in time) evolution with another one.

The modulus squared of the fidelity amplitude $F(t)= \vert f(t) \vert^2$ is
the fidelity, a standard benchmark in quantum information \cite{NC00a,
libroberman}. Note that, although we do not mark this explicitly, $f$ and $F$
are dependent on the choice of the initial state $|\psi\>$. Often the average
fidelity over some set of states, typically randomly chosen, is used.  The
importance of this measure resides in the fact that it has minimal bias, as
it averages errors over the entire Hilbert space. Other choices are rather
task-dependent, but certainly important. Thus e.g. the success rate for a
given task is, in that sense, the most practical.

After this brief introduction to fidelity decay we shall
present the random matrix model we use, and then discuss
its solution by linear response. Next we shall compare exact solutions
 to the approximate ones we obtained. We shall then mention the concept of
scattering fidelity to be able to compare theory to experiments from well
outside quantum information.

\subsection{The random matrix model} \label{sec:RMTforfidelity}

The model describes fidelity decay in a quantum-chaotic (\ie mixing) system,
suffering a global static perturbation; that model could be extended to treat
also noisy perturbations. Typically such a perturbation is less harmful, and
can furthermore be treated using the statistical properties of the noise
directly in the correlation functions \cite{reflosch}.  Chaoticity justifies
choosing the unperturbed system from one of the classical ensembles or more
particularly from a GUE.  The word ``global'' implies that, in the eigenbasis
of the unperturbed Hamiltonian, the perturbation matrix must not be sparse.

We consider a perturbed Hamiltonian of the form
\begin{equation}
H_{\eps'} = \cos(\eps ')\; H_0 + \sin(\eps ')\; H_1 \: ,
\label{R:Hcos}
\end{equation}
where $H_0$ and $H_1$ are chosen randomly from one of the classical
ensembles.  Using this form rather then the more conventional $H_{\epsilon '}
= H_0 + \epsilon' H_1 $ has the advantage that the perturbation does not
change the level density of the Hamiltonian. Changing the level density is a
rather trivial form of perturbation; even if we only scale the spectrum we
will have fidelity decay due to dephasing in its most primitive form. When
discussing the experiments, we shall return to this point. While the use of
the classical ensembles is fully justified for $H_0$ (by the type of system
we discuss), this is not obviously true for the perturbation.  Non-global
perturbations may occur and behave non-generically \cite{Stockk-localpert}
but also global perturbations with different properties could be important,
such as may result two-body interactions \cite{pizorn:035122} or from
maximally time-reversal breaking, \ie hermitian antisymmetric perturbations
of a symmetric (GOE) Hamiltonian \cite{gorin:244105}.  We shall discuss such
situations at the end of this section.

We are interested in situations where $\eps'$ scales as $1/\sqrt{N}$, where
$N$ denotes the dimension of the Hamiltonian matrices.  The matrix elements
of the perturbation then couple a finite number of neighboring eigenstates of
the unperturbed system, largely independent on $N$.  Stronger perturbations
would practically lead to a loss of fidelity on a time scale of the order of
the Zenon time given by the inverse of the spectral span of the Hamiltonian.
Considering large $N$, we linearize the trigonometric functions in
\eref{R:Hcos}. We furthermore fix the average level spacing of $H_0$ to be
one in the center of the spectrum, and require that the off-diagonal matrix
elements of $V= H_1/\sqrt{N}$ have unit variance. This leads us to the
conventional form
\begin{equation}
H_\eps = H_0 + \eps\; V \; ,
\label{R:Heps}\end{equation}
if we rescale the perturbation parameter as $\eps = \sqrt{N}\eps'$.
It is easy to check that corrections to the Heisenberg time are of
order $\mathcal{O}(1/N)$. We may use different ensembles for $H_0$ and $V$.
In many cases, the ensemble of perturbations is invariant under the
transformations that diagonalize $H_0$. We can then choose $H_0$ to be
diagonal and with a spectrum $\{E_j\}$.  In this situation we can unfold the
spectrum that defines $H_0$, to have average level density one along the
entire spectrum, or we can restrict our considerations to the center of the
spectrum. This restricts us to situations, where the spectral density may be
assumed constant over the energy spread of the initial state. Other cases
could be important but have, to our knowledge, so far not been considered in
RMT.

In this section, the eigenbasis of $H_0$ will be the only preferred basis in
contrast to the following section, where we deal with entangled subsystems.
Unless stated otherwise, we consider initial states to be random, but of
finite span in the spectrum of $H_0$. The spectral span of the initial state
and the spreading width of the $H_0$-eigenstates, in the eigenbasis of
$H_\eps$, determine the only additional relevant time scales. They should be compared to
the Heisenberg time $\tau_\rmH$, by unfolding the spectrum of $H_0$. In the
limit $N\to\infty$, the Zeno time (of order $\tau_\rmH/N$) plays no role.

The results presented here cover essentially the range from the perturbative
up to the Fermi golden rule regime~\cite{Jacquod:01, Cerruti:02}.  The
analysis of the quantum freeze and an exact analytical result for the random
matrix model will provide additional and/or different regimes.  The Lyapunov
regime~\cite{Jacquod:01, Jalabert:01} as well as the particular behavior of
coherent states are certainly not within the scope of RMT.

Finally, we wish to add that, in many situations, the fidelity amplitude $f$
is self averaging [\eref{Q2LR:F}]. Therefore we mainly concentrate on the
fidelity amplitude, and do not bother with the more complicated averages for
fidelity $F$ itself .

\subsection{Linear response theory and RMT}\label{Q2LR} 

We shall follow the approach of~\cite{GPS04}, which uses the linear response
approximation expressed in terms of correlation integrals, as introduced by
Prosen \cite{0305-4470-35-6-309} and discussed in \cite{reflosch}.  Some
preliminaries about units and a brief recapitulation of the interaction
picture will be useful.

First we obtain a particularly useful expansion of the echo operator
\eref{eq:defEcho}, namely the Born expansion. The
definition of the state ket in the interaction picture is
\begin{equation}
 \label{eq:defstateinteract}
 |\psi(t)\>_I = U_0^\dagger (t) U_\eps (t)|\psi(0)\>=M_\eps (t)|\psi(0)\>
\end{equation}
where $U_\eps (t)=\exp(-\rmi t  H_\eps)$ is the evolution
operator corresponding to Hamiltonian (\ref{R:Heps}).
Consider the equation of
motion of the ket in the interaction picture ($\hbar=1$):
\begin{equation}
 \label{eq:motioninteract}
 \rmi \frac{\rmd |\psi(t)\>_I}{\rmd t}= \eps \tilde{V}_t |\psi(t)\>_I
\end{equation}
where we are using the shorthand
\begin{equation}\label{eq:deftilde}
\tilde A_t= U_0^\dagger(t) A U_0(t)
\end{equation}
for any operator $A$ in the interaction picture.
Formal integration of \eref{eq:motioninteract} leads to
\begin{equation}
 \label{eq:formalintegrationbornone}
 |\psi(t)\>_I=|\psi(0)\>-
   \rmi \eps \int_0^t \rmd \tau \tilde V_\tau |\psi(\tau)\>_I.
\end{equation}
Solving the integral by iteration we obtain
\begin{equation}
 \label{eq:formalintegrationborn}
 |\psi(t)\>_I=\left( \openone
   -\rmi\eps \int_0^t \rmd \tau \tilde V_\tau
   - \eps^2 \int_0^t \rmd \tau \int_0^\tau \rmd \tau' \tilde V_{\tau'}
   + \ldots \right) |\psi(0)\>.
\end{equation}
Comparing \eqref{eq:defstateinteract}  and \eqref{eq:formalintegrationborn} we
obtain the Born expansion:
\begin{equation}
 \label{eq:bornexpansionAPP}
 M_\eps (t)= \openone -\rmi \eps \int_0^t \rmd \tau \tilde V_\tau
   - \eps^2 \int_0^t \rmd \tau \int_0^\tau \rmd \tau'  \tilde V_\tau \tilde V_{\tau'}
   + \Or(\eps^3).
\end{equation}
From \eref{eq:definefidelityecho} we see that we must only calculate the
expectation value of the echo operator, and thus of the two objects
$\int \<  \tilde V_\tau \>$ and $\iint \< \tilde V_\tau \tilde V_{\tau'}\>$.
At this point we choose to work in the eigenbasis of the unperturbed Hamiltonian
where we can write
\begin{equation}
H_0 = {\rm diag}(E_\alpha),  \quad U_0(t) =
{\rm diag}(\rme^{-\rmi E_\alpha t/\hbar}) \; .
\end{equation}
We proceed averaging the perturbation over the GUE. This average
will be denoted by $\la\cdot\ra_V$. As the matrix elements
are independent (up to symmetry requirements) complex Gaussian variables
we have that
\begin{equation}
\la V_{i,j}\ra_V=0,\quad
\la V_{i,j} V_{k,l}\ra_V = \delta_{i,l}\delta_{j,k}.
\label{AL_GEdef}
\end{equation}
This relation fixes a normalization condition on the ensemble. However
this normalization condition can always be met, absorbing the appropriate
factor in $\eps$.

The expectation value of the linear term of the echo operator yields
zero automatically, as we can check for the matrix element 
$[\cdot]_{\alpha, \beta}$:
\begin{equation}
\left[ \left\langle \int_0^t \rmd \tau
     \tilde V_\tau \right\rangle_V \right]_{\alpha,\beta}=
 \int_0^t \rmd\tau  \rme^{\rmi \tau (E_\alpha - E_\beta)}
   \la V_{\alpha,\beta}\ra_V = 0.
\end{equation}
The next term in the expansion involves the correlation function $\la\tilde
V_\tau \tilde V_{\tau'}\ra_V$.  We obtain:
\begin{align}
\la [\tilde V_\tau \tilde V_{\tau'}]_{\alpha,\beta}\ra_V
 &= \sum_\gamma  \la \rme^{\rmi E_\alpha \tau}  V_{\alpha,\gamma}
\rme^{\rmi E_\gamma (\tau' - \tau)} V_{\gamma,\beta} \rme^{\rmi E_\beta \tau'}
\ra_V
 \nonumber\\
&= \sum_\gamma
\rme^{\rmi \tau (E_\alpha - E_\gamma) }
\rme^{\rmi \tau' (E_\gamma - E_\beta) }
  \delta_{\alpha,\beta}\nonumber\\
&= \delta_{\alpha,\beta} \sum_\gamma \rme^{\rmi(E_\gamma -E_\alpha)\, ( \tau-\tau')}  .
\label{calC}\end{align}
We now average over the initial state. 
Consider an observable $A$, and the state $|\psi\>=\sum_i x_i |i\>$,
with $x_i$ complex random numbers with standard deviation $1/N$ (for
normalization), \ie a random state.
Then, denoting average over the initial state as
$\la\cdot\ra_{|\psi_\>}$,
\begin{equation}
 \left\la  \<\psi |A|\psi\> \right\ra_{|\psi\>}=
\sum_{i,j} A_{i,j} \left\la x_i^* x_j\right\ra_{|\psi_\>}
=\sum_{i,j}A_{i,j} \frac{1}{N} \delta_{i,j} =\frac{1}{N}\tr A.
\label{eq:averageistrace}
\end{equation}
In other words,  to average expectation values
over random states is equivalent to tracing the operator.
Thus, after evaluating the average over initial conditions of the
correlation function, we obtain
\begin{equation}
 \left\la
 \tilde V_\tau \tilde V_{\tau'}
 \right\ra_{V,|\psi\>}=
\frac{1}{N} \tr \sum_\gamma \rme^{\rmi(E_\gamma -E_\alpha)\, ( \tau-\tau')}
     |\alpha\>\<\alpha |=
\frac{1}{N} \sum_{\alpha,\gamma} \rme^{\rmi(E_\gamma -E_\alpha)\, ( \tau-\tau')}
=K_2 (\tau-\tau').
\label{eq:toformfactor}
\end{equation}
We have thus related fidelity decay {\it directly} to the
form factor of the environment. For the average fidelity amplitude
we obtain 
\begin{multline}
 \< f(t) \>= 
   1-\eps^2\int_0^t \rmd \tau \int_0^\tau \rmd \tau' K_2 (\tau-\tau')\\
 =1-\eps^2 \left[
 \frac{t \tau_\rmH}{2} + \frac{t^2}{2} - \int_0^t \rmd \tau \int_0^\tau \rmd \tau'
              b_2\left( \frac{\tau-\tau'}{\tau_\rmH} \right) \right].
\label{eq:fidelityGUE}
\end{multline}
Any stationary ensemble from which $H_0$ may be chosen, yields a particular
two-point function $b_2$. The correlation integral over $b_2$ for the GOE
is discussed in~\cite{GPS04}. If we use a PUE, \ie a random level sequence,
$b_2(t)=0$, the last term in~(\ref{eq:fidelityGUE}) vanishes. Typically
(at least in the case of the classical ensembles), spectral correlations lead
to a positive $b_2$, such that fidelity decay will be slowed down by these
correlations.  However the dominant term is, before/after the Heisenberg time,
the linear/quadratic one respectively. The correlation integral for the GUE is
given by
\begin{equation}
\int_0^t \rmd \tau \int_0^\tau \rmd \tau'
              b_2\left( \frac{\tau-\tau'}{\tau_\rmH} \right)=
              t\min \{t,\tau_\rmH\}-\frac{\min\{t,\tau_\rmH\}^3}{3\tau_\rmH}.
\label{eq:dI}
\end{equation}
Averaging the perturbation over other ensembles is also possible. In the
final result \eref{eq:fidelityGUE}, on has to substitute $t^2/2$ with
$t^2/\beta_V$ where $\beta_V$ labels the ensemble from which $V$ is drawn
\cite{GPS04}.  The result~(\ref{eq:fidelityGUE}) shows two remarkable
features: The first is that the linear and the quadratic term in $t$ both
scale with $\eps^2$.  The second is about the role of the two possibly
different ensembles used for the perturbation and for $H_0$. As long as the
invariance group of the perturbation allows to diagonalize $H_0$ the
characteristics of $V$ affect only the prefactor of the $t^2$-term, while the
characteristics of $H_0$ affect only the two-point form factor $b_2$.

The expansion in time, \eref{eq:fidelityGUE}, contains the leading terms for
both the regime known as perturbative and the one known as the Fermi golden
rule regime~\cite{Jacquod:01}.  Both are exponentials of the corresponding
terms in the linear response approximation. It is then tempting to simply
exponentiate the entire $\eps^2$-term to obtain
\begin{equation}
\la f(t)\ra = \exp\left\{-\eps^2 \left[
  \frac{t \tau_\rmH}{2}+ \frac{t^2}{\beta_V}
  - \int_0^t\rmd\tau\int_0^{\tau}\rmd\tau'
      b_2\left(\frac{\tau-\tau'}{\tau_\rmH}\right)
      \right]\right\}
\label{Q2LR:expLR} \end{equation}
This expression will prove to be extremely accurate for perturbation
strengths up to the Fermi golden rule regime. Some justification for the
exponentiation is given in~\cite{Prosen:03ptps}. While exponentiation in the
perturbative regime is trivially justified, our result shows that for times
$t\ll \tau_\rmH$, we always need the linear term in $t$ to obtain the correct
answer; with other words, the separation into the two regimes is somewhat
artificial: Before the Heisenberg time, the linear term is dominant, and
except if fidelity is way down before Heisenberg time, after this time the
quadratic term will dominate. Thus the RMT
formula is adequate to describe the entire transition between the
perturbative and the Fermi golden rule regime. In experiments the interplay
of both terms is important~\cite{gorin-weaver,SGSS05,expRudi} as we shall see
below. On the other hand, for stronger perturbations fidelity has decayed
before Heisenberg time to levels where our
approximation fails.  Comparison with the exact
result~\cite{1367-2630-6-1-199,StoSch04} will show that the exponentiation
allows to extend the linear response result from a validity of $\la f(t)\ra
\approx 1$ to a validity range of $\la f(t)\ra \gtrsim 0.1$

Note that the pure linear response result is probably all we need for quantum
information purposes, as processes with fidelity less than $1-\eta$, where
$\eta \sim 10^{-2}$, are not amenable to quantum error correction
schemes~\cite{NC00a}. The exact treatment will show where to expect
additional effects, but experiments at this time are still limited to $\la
f(t)\ra \gtrsim 0.1$~\cite{SGSS05,expRudi,gorin-weaver}.

We now turn our attention to fidelity.  It can be calculated in the linear
response approximation along the same lines as above. One
obtains~\cite{GPS04}:
\begin{equation}
\la F(t)\ra = \la |f(t)|^2\ra = \la f(t)\ra^2 + \eps^2\;
 (2/\beta_V)\; {\it ipr}\; t^2\; +\mathcal{O}(\eps^4) \; .
\label{Q2LR:F}\end{equation}
Here ${\it ipr} = \sum_{\nu} |\braket{E_\nu}{\psi}|^4$ indicates the {\em
inverse participation ratio} of the initial state expanded in the eigenbasis
of $H_0$. We deduce two extreme effects: On the one hand it shows the
self-averaging properties of this system. For states with a large spectral
span in $H_0$, the correction term that marks the difference between $\la
F(t)\ra$  and $|\la f(t)\ra|^2$ goes to zero as the inverse participation
ratio becomes small ($\sim 1/N$).  On the other hand, for an eigenstate of
$H_0$, ${\it ipr}=1$, and hence the quadratic term in \eref{Q2LR:F}
disappears. Moreover, the correlations cancel the linear term after the
Heisenberg time. Thus, we find that after Heisenberg time the decay stops for
an $H_0$ taken from a GUE and continues only logarithmically for a
GOE~\cite{GPS04} up to the onset of the next term in the Born expansion. 

\subsection{Supersymmetric results for the fidelity amplitude}\label{Q2SS} 

\begin{figure}
\includegraphics[width=7.7cm]{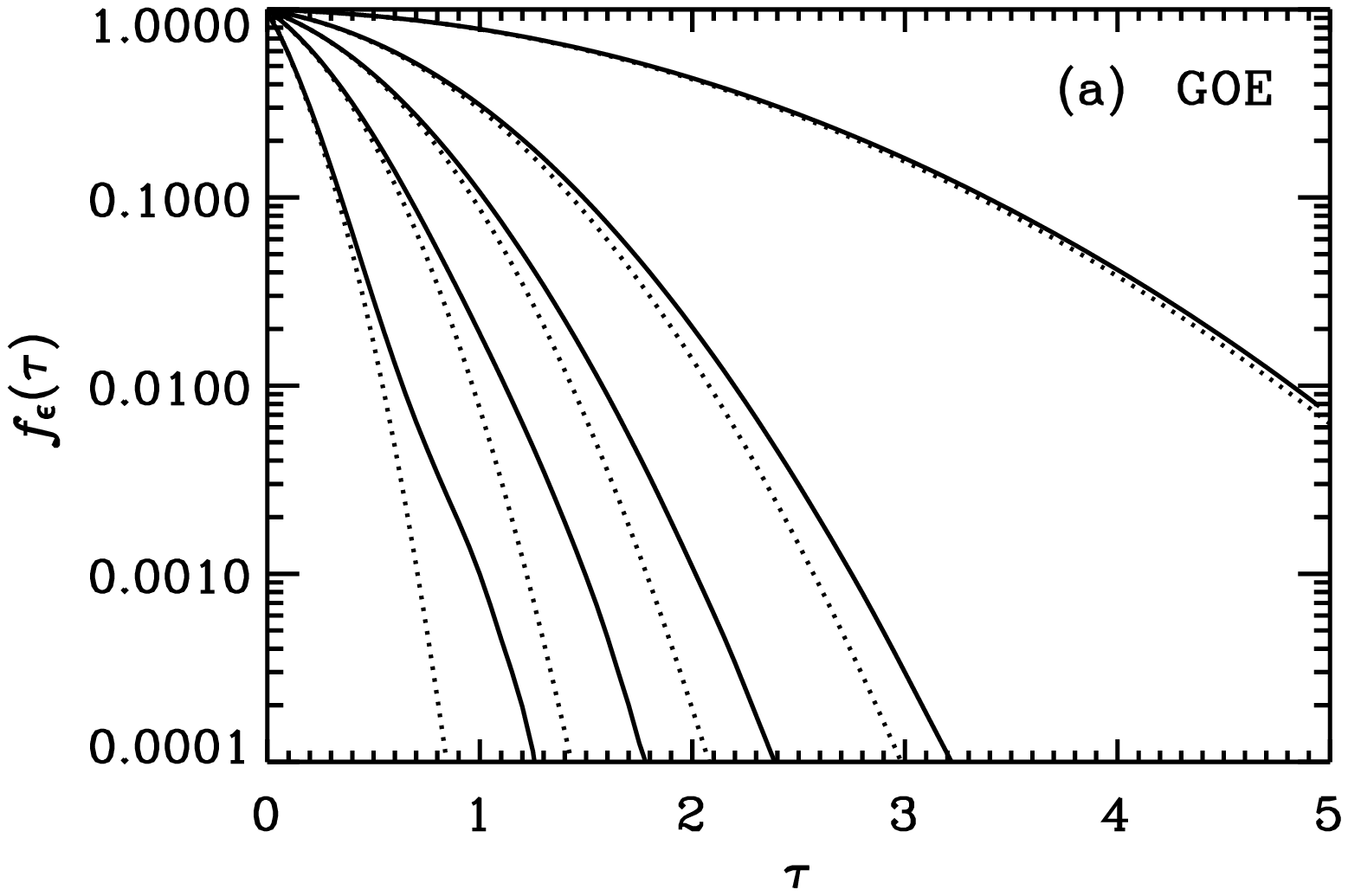}\hfill
\includegraphics[width=7.7cm]{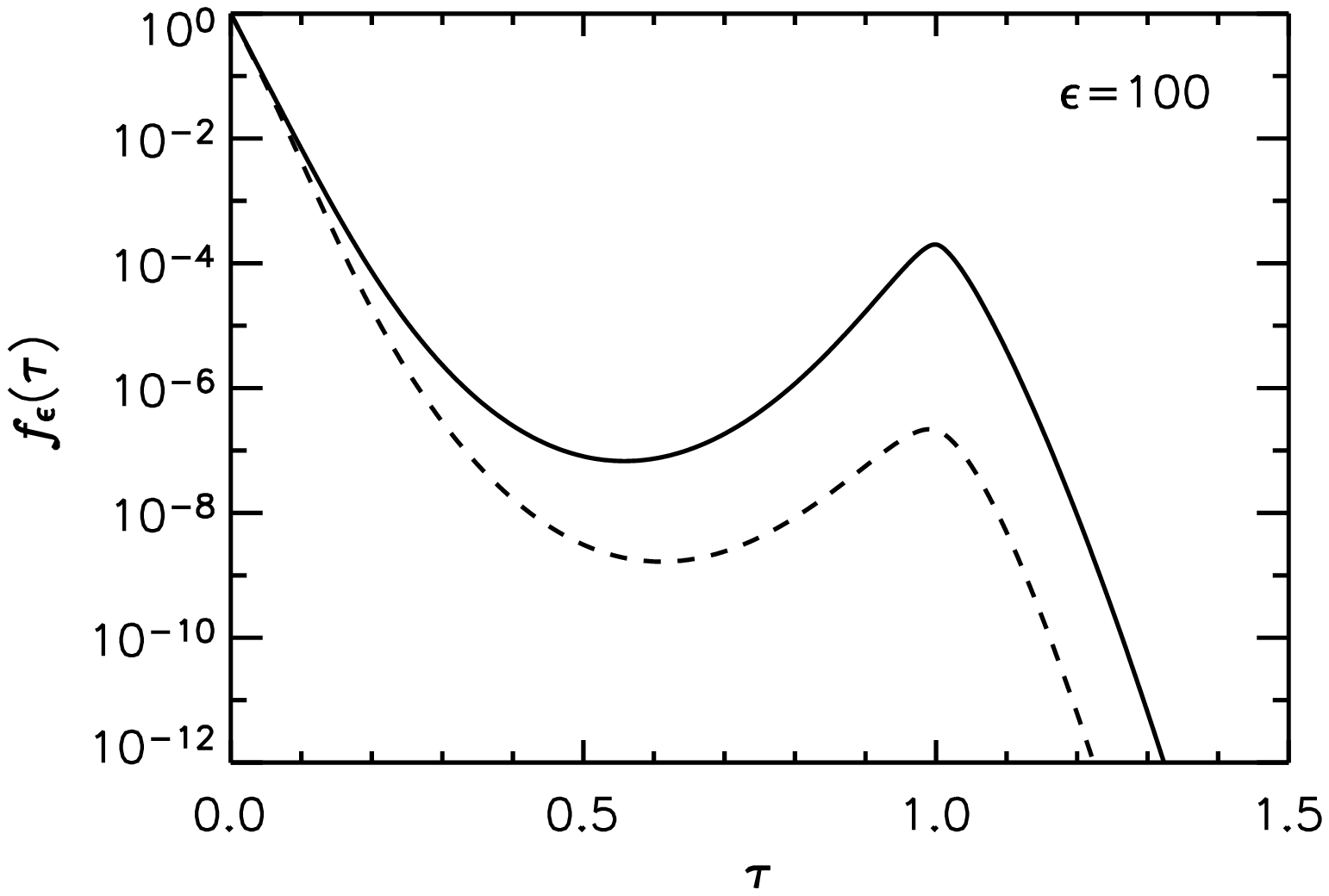}
\caption{Fidelity amplitude decay for the random matrix model, defined in
  \eref{R:Heps} (taken from~\cite{1367-2630-6-1-199}).  Part (a) shows $\la
  f(t)\ra$ for the GOE case, as obtained from the exact expression,
  \eref{Q2SS:GOE}, (solid lines), together with the same quantity, as obtained
  from the exponentiated linear response result, \eref{Q2LR:expLR} (dashed
  lines). The perturbation strength has been set to the following values: $
  \eps^2= 0.2,\, 1\, , 2\, , 4\, $ and $10$.  Part~(b) shows $\la f(t)\ra$
  with $ \eps^2= 100$ for the GUE case (solid line) and the GOE case (dashed
  line), as obtained from the exact expressions, eqs.~(\ref{Q2SS:GUE}) and
  (\ref{Q2SS:GOE}), respectively.  In both figures the authors use the
  convention $\tau=t$ and $f_\eps (\tau)=\la f(t) \ra$.}
\label{Q2SS:f:StoSch}\end{figure}

The exponentiated linear response formula~(\ref{Q2LR:expLR}) agrees very well
with dynamical models~\cite{GPS04,Haug:05} and
experiments~\cite{SGSS05,expRudi,gorin-weaver}.  Yet this
result has to fail for sufficiently strong perturbations, even if we forget
its heuristic justification. Fortunately many problems in RMT can been solved
exactly, and recently St\" ockmann and Sch\"
afer~\cite{1367-2630-6-1-199,StoSch04} have done exactly this for the model
given in \eref{R:Heps}, for GOE or GUE matrices, in the limit of infinite
dimensions. The solution for the GSE will be published shortly
\cite{correlationsRMT}.

More specifically, they choose $H_0$ and $V$ independently but both from the
same classical ensemble, and compute the fidelity amplitude $\la f(t)\ra$
with the help of super-symmetry techniques. For a detailed account we refer to
the original paper~\cite{1367-2630-6-1-199}.  They obtain
\begin{equation}
\la f(t)\ra = \frac{1}{t}\int_0^{{\rm min}(t,1)}\rmd u\;
 (1+t-2u)\; \e^{- \eps^2 (1+t-2u) t/2}
\label{Q2SS:GUE}\end{equation}
for the GUE case and
\begin{align}
\la f(t)\ra &= 2\int_{{\rm max}(0,t-1)}^t\rmd u\int_0^u\rmd v\;
 \frac{(t-u)(1-t+u)\, v\, ( (2u+1)\, t -t^2+v^2)}
 {(t^2-v^2)^2\,\sqrt{(u^2-v^2)((u+1)^2-v^2)}}\notag\\
&\qquad\times \e^{- \eps^2\, [ (2u+1)\, t -t^2+v^2]/2} \; .
\label{Q2SS:GOE}\end{align}
for the GOE case ($\tau_\rmH=1$ in this section). 
These solutions are valid for
arbitrary time independent perturbation strength, in the limit $N\to\infty$.

\begin{figure}
\psfrag{f}{$\log_{10}\la f(t)\ra$}
\psfrag{t}{$t$}
\centerline{\includegraphics[width=0.8\textwidth]{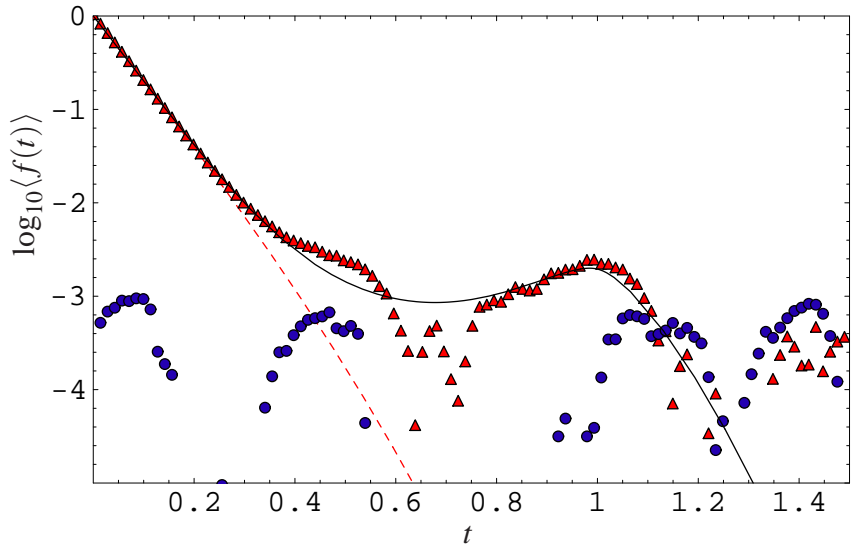}}
\caption{(Color online) The decay of the fidelity amplitude in a dynamical
  system, in the case of broken time-reversal symmetry, in a regime where the
  partial revival is observable (taken from Ref.~\cite{pineda:066120}). More
  precisely, $\log_{10} \la f(t)\ra$ is plotted. The triangles show the real
  part of the fidelity amplitude as obtained from the numerical simulation;
  the circles show the imaginary part, which goes to zero as $1/\sqrt{N}$ due
  to state averaging.  The perturbation strength is $\eps^2= 31.78$. The solid
  line shows the exact theoretical result, \eref{Q2SS:GUE}, the dashed line
  shows the exponentiated linear response result, \eref{Q2LR:expLR}.}
\label{Q2SS:f:pineda}\end{figure}

In Fig.~\ref{Q2SS:f:StoSch}, we reproduce two graphs
from~\cite{1367-2630-6-1-199}.  The left one, compares the exact and the
exponentiated linear response result for $\la f(t)\ra$ for the GOE case.
For large perturbations we find a qualitative difference in the shape of
fidelity decay as a shoulder is forming in the exact results. For even
stronger perturbations, depicted on the right panel, this becomes more
pronounced as a revival is seen at Heisenberg time.  Yet, the revival is
noticeable only for very small fidelities, of the order of $10^{-4}$ for the
GUE and  $10^{-6}$ for the GOE. In all cases agreement with the exponentiated
linear response formula is limited to $\la f(t)\ra \gtrsim 0.1$. Exponentiated linear response
 is thus adequate for most applications. Indeed it was difficult to come up with a
dynamical model which can show the revival. In \cite{pineda:066120} a kicked
Ising spin chain \cite{prosenKI} has been used to illustrate the partial
revival, as shown in Fig.~\ref{Q2SS:f:pineda}.  Taking advantage of the
relative ease of numerical calculations in such a model, the authors used a
multiply kicked Ising spin chain in a Hilbert space spanned by 20 qubits, and
averaged over a few initial conditions. The aim was to obtain the partial
revival with as little averaging as possible, relying on the self-averaging
properties of the fidelity amplitude. The model shows random matrix behavior
up to small deviations as far as its spectral statistics is concerned. The
result for the decay of the fidelity amplitude is reproduced in
Fig.~\ref{Q2SS:f:pineda}.  Yet it will probably be difficult to see this
effect in an experiment.

Thus the random matrix model captures all fidelity decay scenarios from
Gaussian (or perturbative) to exponential (or Fermi-Golden rule). Furthermore
it displays an additional feature, namely the weak revival at Heisenberg time.
Naturally the random matrix model does not capture the Lyapunov regime. This is
an additional semi classical regime for fairly strong perturbations
\cite{Jalabert:01, reflosch} where the decay is determined by the classical
Lyapunov exponent and in some range becomes independent of the perturbation
strength.  It is not obvious that such a regime always exists. For example, for
the kicked spin chain this regime has not been observed, and it may well be
that there is no semi-classical limit for this system.

The revival at Heisenberg time can be related to a revival in the
cross-correlation function of spectra in the time-domain through a recently
discovered direct relation between this cross-correlation and fidelity decay
\cite{correlationsRMT}.

\subsection{A micro wave experiment of fidelity decay}

\begin{figure}
\centerline{ \includegraphics[width=0.48\textwidth]{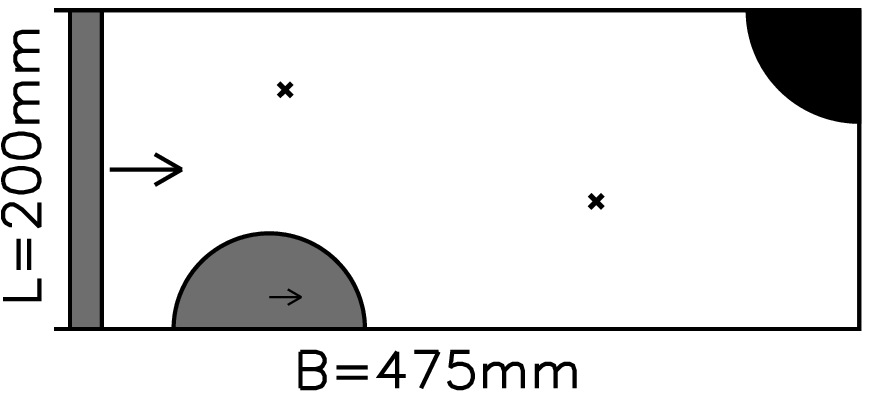}
 \includegraphics[width=0.48\textwidth]{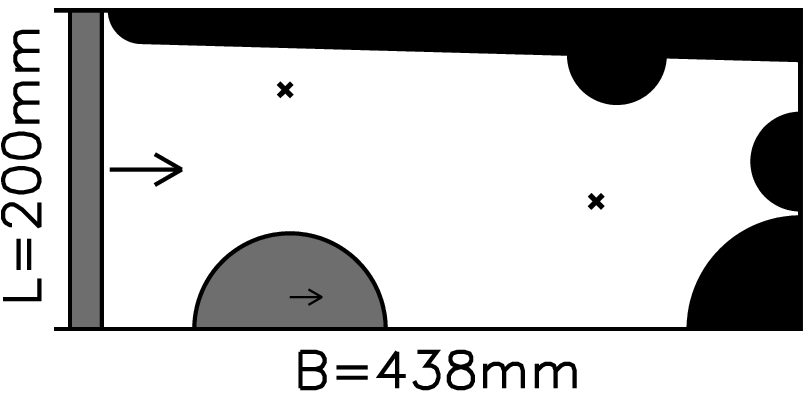}  }
\caption{Geometry of the billiards (figure taken from~\cite{SGSS05}). In the
  billiard in the right figure bouncing ball orbits have been avoided by
  inserting additional elements.}
\label{ES:f:muwavebills}\end{figure}

\begin{figure}
\centerline{
 \includegraphics[width=0.48\textwidth]{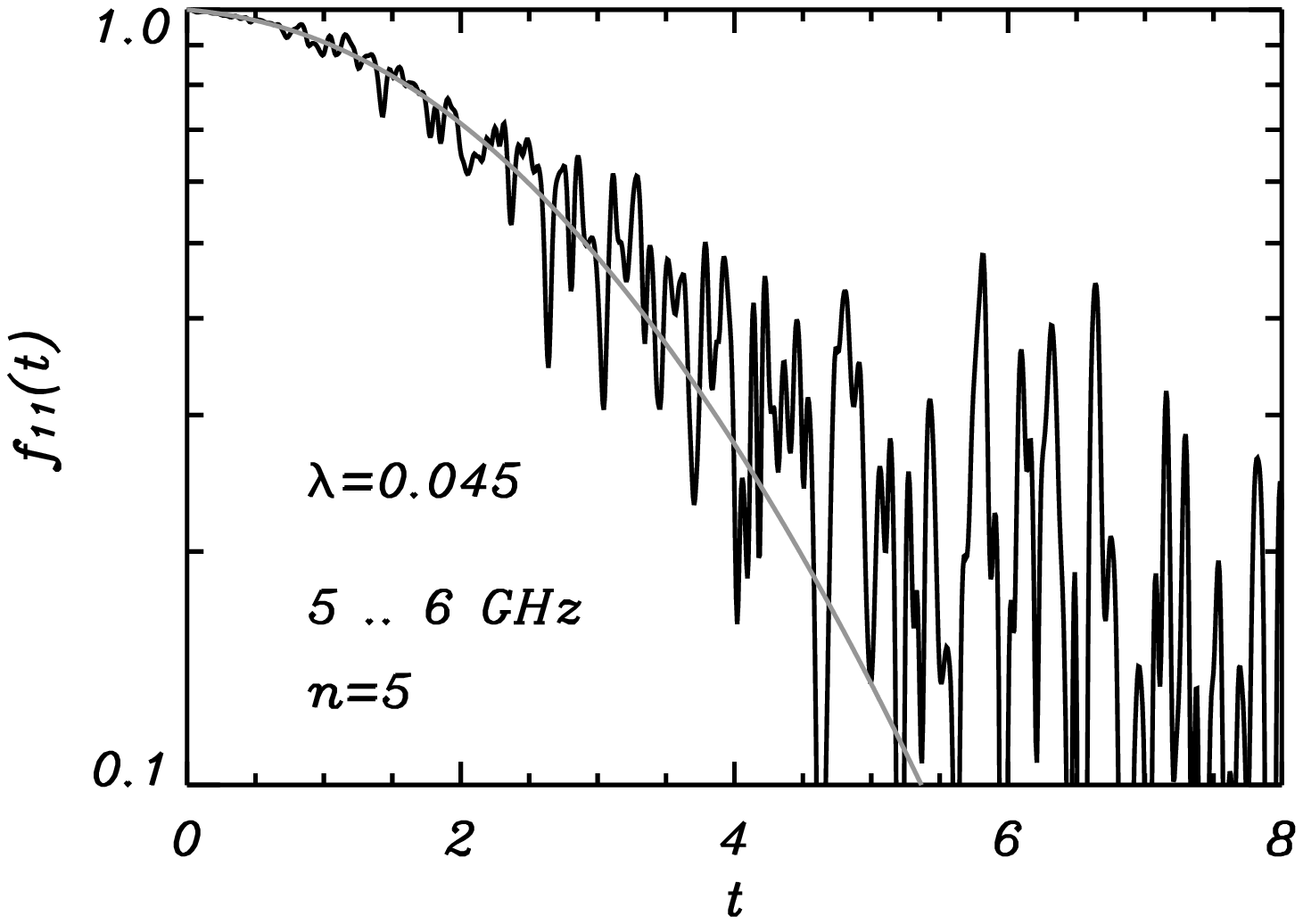}
 \includegraphics[width=0.48\textwidth]{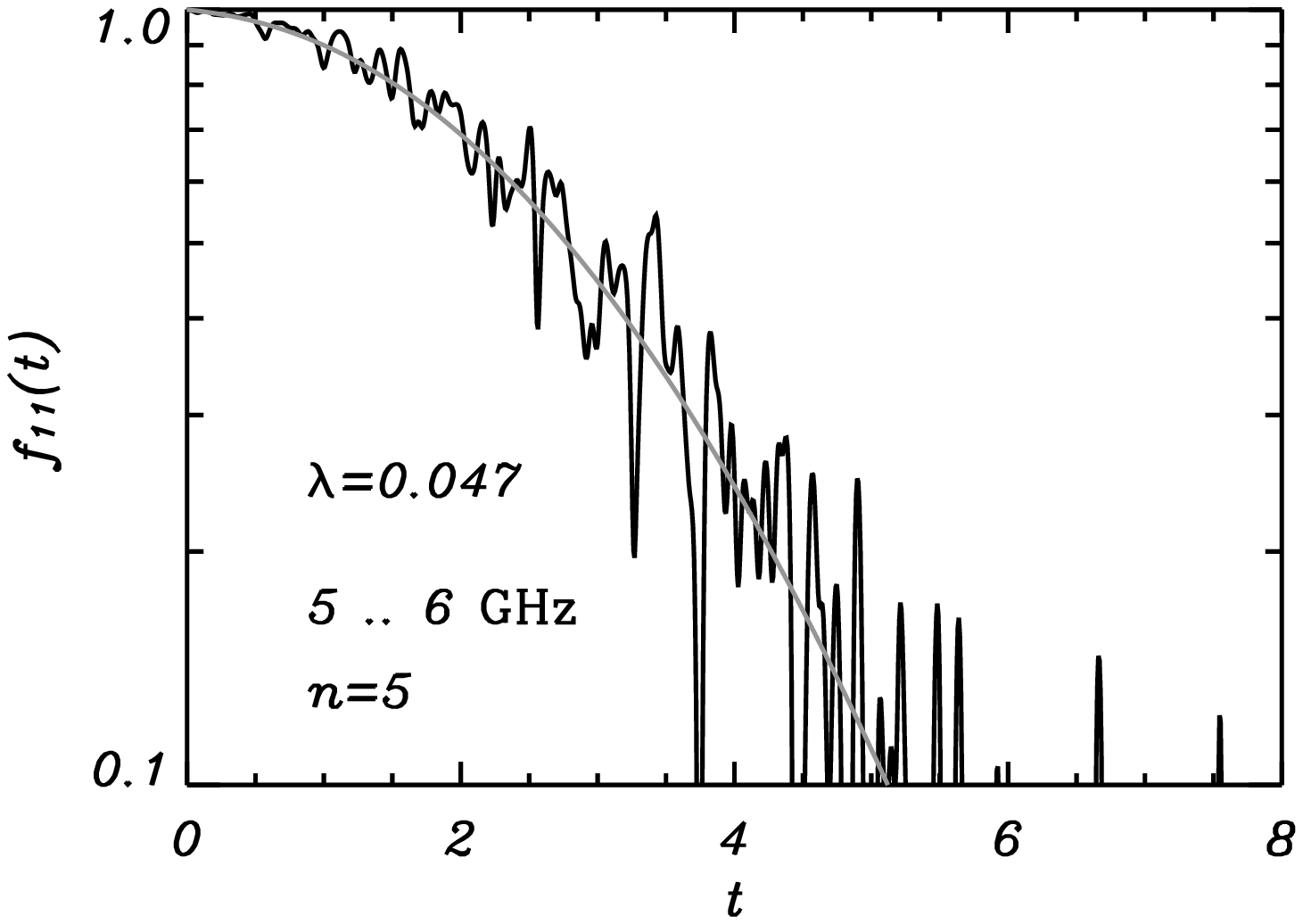} }
\caption{Logarithmic plot of the scattering fidelity $f_{11}(t)$, directly
  related to the fidelity amplitudes (figure
  taken from~\cite{SGSS05}).  The smooth curve shows the linear-response
  result. For the billiard without bouncing balls, the perturbation parameter
  $\eps$ was obtained from the variance of the level velocities; in the other
  case it was fitted to the experimental curve.}
\label{ES:f:mubillcfun-bottom}\end{figure}

The dynamics of classical electromagnetic waves in a thin resonator is
equivalent to the Schr\" odinger equation of a single quantum particle with
two degrees of freedom.  This is exploited in the microwave cavity
experiments, where properties of two dimensional quantum billiards can be
studied. The Marburg group considered  cavities (Fig.~\ref{ES:f:muwavebills})
consisting of a rectangle with inserts that assure chaotic, but not
necessarily hyperbolic, ray behavior.  Both situations, with and without
parabolic manifolds leading to so-called bouncing ball states, were
considered.  The presence of bouncing ball states is known to lead to a weak
deviation from RMT behavior \cite{richterBouncingBall}. This will serve to
check how important small deviation, from the ``universal'' RMT behaviour, can be for
fidelity decay.  One of the walls was movable in small steps allowing the
perturbation $V$ occurring in an echo experiment [see \eref{R:Heps}]. It is
important to note that the shift of the wall changes the mean level density
and by consequence the Heisenberg time.  This trivial perturbation, which
would cause very rapid fidelity decay, is eliminated in this case by
measuring all times in proper dimensionless units, \ie in terms of the
Heisenberg time. Two antennae allow to measure both reflection and
transmission channels.  The experiment was carried out in the frequency
domain. Afterwards the Fourier transform is used to obtain correlation
functions in the time domain.  We thus get autocorrelation functions for any
given configuration, as well as cross correlation functions associated with
the different positions of the moving wall, \ie between evolutions with
different Hamiltonians.  After normalizing the latter with the
autocorrelation we find the so called scattering fidelity.  For chaotic
systems, in the weak perturbation limit, scattering fidelity has been shown
to be equivalent to the fidelity amplitude \cite{SGSS05}.  Figure
\ref{ES:f:mubillcfun-bottom} shows the experimental and theoretical result in
excellent agreement in the absence of bouncing ball states and, as expected,
in lesser agreement if they are present.  In principle in our model the
perturbation strength $\epsilon$ is a free parameter. Yet in this case it was
obtained independently by exploring the level-dynamics, \ie the movement of
the energy levels under change of the Hamiltonian.  The same random matrix
model can account for this situation.  For low frequency ranges where
resonances are well separated, this so-called level dynamics can be measured.
The perturbation strength was extracted and used in the formula for fidelity
decay.

Similar experiments have been performed by Weaver and Lobkis \cite{LobWea03}
on the coda of vibrations of metal blocks. Using the concept of scattering
fidelity, these experiments can be reinterpreted as fidelity measurements
\cite{gorin-weaver}.  Quantum optics techniques in principle allow direct
measurement of the fidelity amplitude, using a single qubit as a probe. The
general context in which this is possible is shown in \cite{GPSS04} and a
detailed proposal for the experiment is given in \cite{Haug:05} for the case
of a kicked rotor where the kicking strength is perturbed.  Actual
experiments where not performed yet, though an experiment with four different
Hamiltonians, two for the forward and two for the backward evolution was
performed \cite{AndDav03}.

\subsection{Residual perturbations and the Quantum freeze}\label{Q2QF} 

After the Heisenberg time, fidelity decay is essentially Gaussian, if it has
not yet decayed significantly [see \eref{Q2LR:expLR}].  This decay is
determined by the diagonal elements of the perturbation (in the interaction
picture) in the eigenbasis of $H_0$.  If this term is zero or very small, we
should see a considerable slowing down of fidelity decay.  That such a
possibility exists was first noted for particular integrable \cite{Prosen:03}
and chaotic \cite{Prosen:05} dynamical systems.  Yet the generality of the
phenomenon is best understood in the context of the RMT model discussed in
\cite{reflosch, gorin:244105}.

Sticking to the representation where $H_0$ is diagonal, we shall thus
consider situations where the perturbation is some random matrix with zero
diagonal. The off diagonal elements will form the so called residual
interaction $V_{\rm res}=V$. Many ways to implement such a residual
perturbation are conceivable. Most are basis dependent. The case, where $H_0$
is chosen from a GOE and $V$ is an antisymmetric Hermitian random Gaussian
matrix is of interest, because the real matrix $H_0$ can be diagonalized
without disturbing the residual (\ie zero diagonal) character of the
perturbation. Thus it has strong invariance properties and indeed this case
is the only random matrix model, so far, for which an exact analytical result
has been obtained~\cite{gorin:244105,Heiner06} by supersymmetric techniques.

\begin{figure}
\centering
\includegraphics{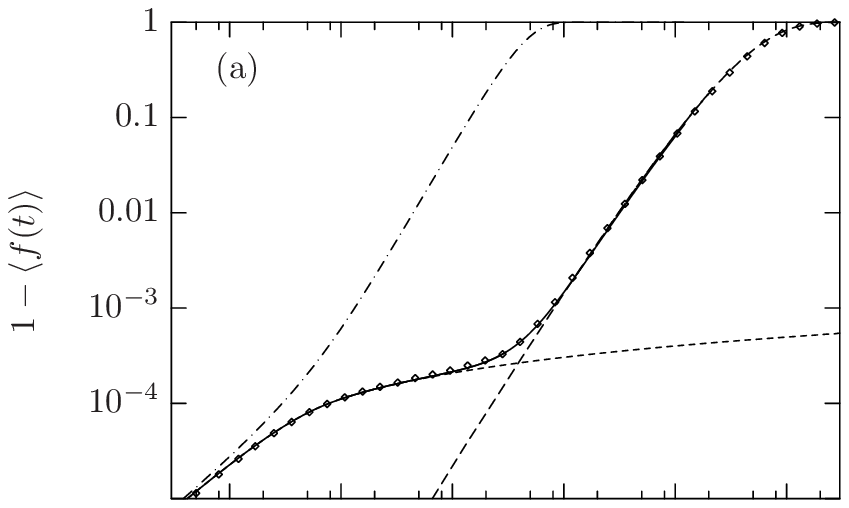} \hspace{-9.6cm} \raisebox{-5cm}{\includegraphics{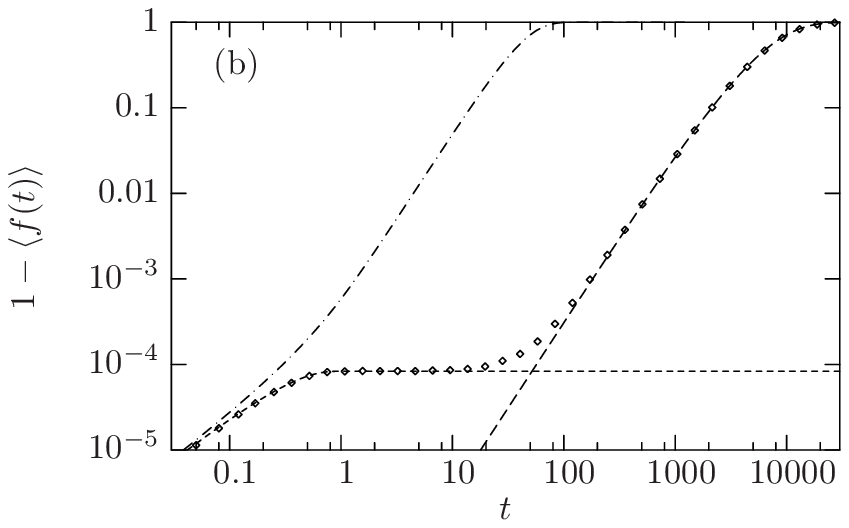}}
\caption{The complement of the average fidelity amplitude $1-\la f(t)\ra$, for
  a weak perturbation, $\eps= 5\times 10^{-5}$.  Frame (a) shows the GOE case
  with a purely imaginary antisymmetric random Gaussian perturbation.  Frame
  (b) shows the GUE case, with an independent GUE with deleted diagonal as
  perturbation.  The exponentiated linear response approximations are plotted
  with short dashed lines (for comparison, the exact results for full GOE and
  GUE perturbations are shown with chain curves).  The forth order results
  \cite{SGSS05} from time-independent perturbation theory are plotted with long
  dashed lines. In the GOE case, the exact analytical result is plotted with a
  solid line. The random matrix simulations are plotted with points. (Taken
  from \cite{gorin:244105}.) }
\label{fig:prlFreezeRMT} \end{figure}

Consider the case of a random perturbation (of GUE or GOE type) with diagonal
elements set to zero.  We have a clearcut stabilization of fidelity after the
Heisenberg time because the double integral in the linear response expression
cancels the linear term in $t$ exactly for the GUE case and (up to a
logarithmic correction) the GOE case.  As mentioned, above the quadratic
term is absent, and decay, as obtained by linear response, essentially ends
at the Heisenberg time.  Decay at much larger times results from the next
term in the Born series and will go as $\epsilon ^4$.

We see in Figure \ref{fig:prlFreezeRMT} that Monte Carlo calculations agree
perfectly with the exact solution for the GOE case with antisymmetric
perturbation. Furthermore both the GOE and the GUE cases are well described
by the linear response result up to the plateau and the end of the plateau
agrees with the behavior of $\eps ^4$ term.  Experiments are not available
yet, but good agreement was shown with a dynamical model \cite{reflosch,
gorin:244105}.

The case with zeros forced on the diagonal of the perturbation is not as
artificial as it might seem, because is is a standard practice of mean field
theory to include the diagonal part of the residual interaction, which does
not affect eigenfunctions, in the zero order mean field Hamiltonian. This
leaves the residual interaction with zero diagonal. Unfortunately, in such a
situation, even if the single particle spectrum is of RMT type, the spectrum
of the $n$-particle problem without residual interaction looks more like a
random spectrum, and this gets worth if the diagonal term of the interaction
is included in $H_0$. In such a case the linear term after Heisenberg time is
not canceled, and we find linear decay after Heisenberg time \cite{reflosch}.
In \cite{pizorn:035122} fidelity decay of two-body random ensembles for $n$
Fermions were studied including the diagonal part of the two-body interaction
in $H_0$. The linear decay for the average fidelity decay was found.  Yet
remarkably it was shown, that the median behavior does display the freeze.
This means, that in a typical realization the freeze is present and thus the
time evolution of a mean field theory with weak residual interaction will
follow the exact one, after some initial deviation, for a long time.

\section{Purity}

We now turn our attention to the second problem addressed in the
introduction.  We shall study the loss of coherence due to coupling to, and
entanglement with, some outer system. This is what people often refer to as
{\it decoherence}.  Quantifying and eventually controlling the degree of
decoherence is a mayor problem of quantum information.  Assuming a pure state
in a bipartite system $\mcH=\mcH_A \otimes \mcH_B$, one can quantify the
degree of entanglement via the purity.  Let $|\psi\> \in \mcH$ be the pure
state in the whole Hilbert space. The reduced density matrix of system $A$
($B$) is then $\rho_A = \tr_B |\psi\>\<\psi|$ ($\rho_B = \tr_A
|\psi\>\<\psi|$). Purity is simply
\begin{equation}
P=\tr \rho_A^2 =\tr \rho_B^2
\label{eq:defpurity}
\end{equation}
which measures the degree of mixedness of each reduced density matrix. One
could also use other mixedness measures such as the von Neumann entropy. We
prefer purity above others because it has an analytic structure that
simplifies analytic treatment. In particular it is not necessary to
diagonalize the density matrix to obtain purity.

In this section we discuss purity decay, thus quantifying  decoherence. First
we derive a general expression for purity decay of $n$ non interacting
qubits.  We shall obtain a sum rule where each term involves a single
qubit and its entanglement with the remaining ones \cite{GPS-letter}. After
that, we assume random matrix environments and couplings, following the scheme
proposed in \cite{GPS04, pinedaRMTshort} and developed in detail
in \cite{pinedalong}. We shall not develop the model with full generality
to keep the discussion at a basic level, allowing us to go slowly through
the simplest (though representative) cases.

\subsection{$n$ qubit purity decay}

First we derive a general formula for purity decay of $n$
qubits under very general assumptions. The problem is solved using the
enlarged Hilbert space 
\begin{equation} \mcH=\mcH_\rme \otimes \mcH_\rmc \end{equation}
where $\mcH_\rmc$ is the Hilbert space of the central system and $\mcH_\rme$
(with dimension $N_\rme$) is the Hilbert space of the environment. The
central system itself has a structure as it is composed of several qubits:
$\mcH_\rmc=\bigotimes_{i=1}^{n}\mcH_i$ with $\dim \mcH_i=2$.  The Hamiltonian
governing the system has the usual structure
\begin{equation}
H_\lambda= H_\rmc + H_\rme + \lambda V,
\label{eq:generalpurity}
\end{equation}
where $H_\rmc$ acts on the central system, $H_\rme$ on the environment and
$\lambda V$ is the coupling between central system and environment.  We shall
moreover require that the Hamiltonian of the central system is local, \ie
$H_\rmc=\sum_{i=1}^{n} H_i$ where $H_i$ acts on $\mcH_i$. We further require
that each qubit is separately coupled to the environment:
\begin{equation}
\lambda V=\sum_{i=1}^{n} \lambda_i V^{(i)}
\label{eq:Vdecos}
\end{equation}
with $V^{(i)}$ acting on $\mcH_\rme\otimes\mcH_i$. We do not
restrict any component of the Hamiltonian to be time independent,
though we shall not explicitly show the time orderings required if
such dependence exists.

The state in the whole Hilbert space after a time $t$ is
$|\psi(t)\>=U_\lambda (t) |\psi_0\>$. $U_\lambda (t)$ is
the evolution operator at time $t$ associated with the Hamiltonian
(\ref{eq:generalpurity}) and $|\psi_0\>=|\psi_\rmc\>\otimes|\psi_\rme\>$
($|\psi_\rmc\> \in \mcH_\rmc$, $|\psi_\rme\> \in \mcH_\rme$)
is a separable pure initial state.
We write this state as $|\psi_0\>=\sum_\mu x_\mu |\mu\>$ or,
using tensor notation, $|\psi_0\>=\sum_{i,j,k} x_{ijk} |ijk\>$ with
$|ijk\>$ being an element of an orthonormal separable basis in
$\mcH=\mcH_\rme \otimes \mcH_\rmq \otimes \mcH_\rms$. We shall keep the convention that
Greek indices are used in the whole Hilbert space whereas Latin ones
are used for subsystems.

Purity is the trace of the square of the density matrix of
the system in question [\eref{eq:defpurity}]. Thus, we are interested in calculating
\begin{equation}
P(t)
= \tr_\rmc \left( \tr_\rme |\psi(t)\> \<\psi(t) | \right)^2
= \tr_\rmc \left[ \tr_\rme M_\lambda(t)
     |\psi_0\> \<\psi_0 | M^\dagger_\lambda(t) \right]^2
\end{equation}
where $M_\lambda(t)$ is the echo operator defined in conjunction with
fidelity, see section \ref{R}.  The last equality is obtained noticing that
the difference between the forward evolution operator $U_\lambda(t)$ and the
echo operator $M_\lambda(t)$ is the local (with respect to
environment-central system separation) operation
$U_0^\dagger(t)$. Since entanglement
properties remain unchanged under local operations, purity has the same value
for the state evolved with either the forward or the echo evolution
operator. Since for large purities  $M_\lambda(t)\approx \openone$,
a series expansion is again feasible for long times. We obtain
\begin{equation}   \label{eq:PurityAfterBorn}
P(t) =1-2\lambda^2 \into \rmRe A(\tau,\tau')+\Or\left(\lambda^4\right),
\end{equation}
where $A(\tau,\tau')=A_J(\tau,\tau')-A_1(\tau,\tau') + A_2(\tau,\tau')-A_3(\tau,\tau')$
and
\begin{subequations}\label{eq:lasas}
\begin{align}
A_J(\tau,\tau')&= p[\tilde V_\tau \tilde V_{\tau'} \varrho_0\otimes\varrho_0]
  = x_\mu x_{i'jk}^* x_{i'j'k'} x_{ij'k'}^*
       [\tilde V_\tau \tilde V_{\tau'}]_{ijk,\mu},   \\
A_2(\tau,\tau')&= p[\tilde V_{\tau'} \varrho_0 \tilde V_\tau \otimes\varrho_0]
  = x_\mu x_\nu^* x_{i'j'k'} x_{ij'k'}^*
    [\tilde V_\tau]_{ijk,\mu} [\tilde V_{\tau'}]_{i'jk,\nu}^*,\\
A_3(\tau,\tau')&= p[\tilde V_{\tau'} \varrho_0\otimes \varrho_0 \tilde V_\tau]
  = x_\mu x_{i'jk}^* x_{i'j'k'} x_\nu^*
    [\tilde V_\tau]_{ijk,\mu}  [\tilde V_{\tau'}]_{ij'k',\nu}^*
\end{align}
\end{subequations}
(summation over repeated indices is assumed in these equations).
The  bilinear  functional
$p [ \rho_1 \otimes \rho_2]= \tr( \tr_\rme  \rho_1 \tr_\rme \rho_2)$
has been introduced, together with $\rho_0=|\psi_0\>\<\psi_0|$,
to simplify the expressions. Notice that each term depends on the coupling in
the interaction picture, see \eref{eq:deftilde}.

Taking advantage of the particular structure of the coupling [\eref{eq:Vdecos}]
we can rewrite the integrand in \eref{eq:PurityAfterBorn}
as a sum of terms, each with two indices labeling the qubits.
$$\lambda^2 A(\tau,\tau')=\sum_{i,j=1}^n \lambda_i \lambda_j A^{(i,j)}(\tau,\tau')$$
with
\begin{multline}\label{eq:AIJdecom} A^{(i,j)}(\tau,\tau')= p[\tilde
V^{(i)}_\tau \tilde V^{(j)}_{\tau'} \varrho_0\otimes\varrho_0] - p[\tilde
V^{(i)}_{\tau'} \varrho_0 \tilde V^{(j)}_\tau \otimes\varrho_i]\\ + p[\tilde
V^{(i)}_\tau \varrho_0\otimes \tilde V^{(j)}_{\tau'} \varrho_0] - p[\tilde
V^{(i)}_{\tau'} \varrho_0\otimes \varrho_0 \tilde V^{(j)}_\tau] \; .
\end{multline}
The terms $A^{(i,j)}(\tau,\tau')$ are cross correlation functions for $i \ne j$
and autocorrelation functions for $i=j$. If the cross correlation functions
drop to zero fast enough, these terms can be neglected.  This happens if the
environment Hamiltonian is chaotic \cite{GPS-letter} or if the couplings are
independent from the outset.  Assuming fast decay of cross correlation
functions [$A^{(i\ne j,j)}\approx 0$] we obtain
\begin{align}\label{eq:spectatorpurity}
P(t)&= 1- \sum_{i=1}^n\left( 1- P^{(i)}_{\rm sp}(t) \right),\\
P^{(i)}_{\rm sp}(t)&= 1- 2 \lambda_i^2\into A^{(i,i)} (\tau,\tau') +
\Or\left(\lambda_i^4\right).\nonumber
\end{align}
Each term $P^{(i)}_{\rm sp}(t)$ represents purity decay of {\it the whole} register
when only {\it a single} qubit is coupled to the environment. That configuration
is called {\it spectator configuration}. \Eref{eq:spectatorpurity} is
a sum rule for decoherence and is the central result of \cite{GPS-letter}.

\subsection{RMT in the spectator configuration}

We now solve the problem of a single qubit in the spectator configuration, when
the Hamiltonian of the environment and the coupling are chosen from the
classical ensembles. We use the following  subscripts: ``c'' (for
central system), ``e'' (for environment) , ``q'' (for the coupled
qubit), and ``s'' (for the spectator).

We have seen that the problem can be formulated in terms of the echo operator.
However, in the spectator configuration, the echo operator does not contain the
internal Hamiltonian of the spectator space. Effectively one can thus drop
those terms and write $H_0=H_\rme + H_\rmq$ (with $H_\rmq$ the internal
Hamiltonian of the coupled qubit) and  $H_\lambda =H_0+ \lambda V$ with
$\lambda V $ the coupling between the coupled qubit and the environment. Notice
how no part of the Hamiltonian is acting on the spectator space.

At this point we rederive a result from~\cite{pinedalong}. Purity decay in the
spectator configuration depends only on $\rho_\rmq=\tr_\rms
|\psi_\rmc\>\<\psi_\rmc|$ the reduced density matrix of the coupled qubit alone:
\begin{align}
 P(t)&=\tr_\rmc \left[ \tr_\rme M_\lambda \otimes \openone_\rms |\psi_0\> \<\psi_0|
  M_\lambda^\dagger \otimes \openone_\rms \right]^2\\
   &= \tr_\rme \left[ \tr_\rmc M_\lambda \otimes \openone_\rms |\psi_0\> \<\psi_0|
  M_\lambda^\dagger \otimes \openone_\rms \right]^2\\
  &= \tr_\rme \left[ \tr_\rmq M_\lambda ( \tr_\rms |\psi_0\> \<\psi_0|)
  M_\lambda^\dagger \right]^2\\
  &= \tr_\rme \left[ \tr_\rmq M_\lambda ( \rho_\rmq \otimes |\psi_\rme\> \<\psi_\rme|)
  M_\lambda^\dagger \right]^2.
\end{align}
The first step uses only that the dynamics in the
spectator space is trivial. In the second step we use the property that
for a pure bipartite system, purity does not depend on the subsystem
observed. Thus instead of measuring purity of the quantum register we formally
calculate purity of the environment. In the next step we split the trace over
all the register into the trace over the coupled qubit and the trace over the
spectator qubits. Next, we explicitly shift the trace over the spectator space
to one over the initial state obtaining the final result: purity decay in the
spectator configuration will only depend on the reduced density matrix of
the coupled qubit.

Since to purify a qubit we only need one additional qubit, we can replace
$\mcH_\rms$ by the Hilbert space of {\it a single} qubit thus greatly reducing
the complexity of the problem! We can now refer to the solution published in
\cite{pinedalong} for two qubits without loosing generality. We
carry on the explicit calculations, in a simple case.

We assume that the nontrivial part of $V$ is well described by the GUE, with a
normalization set by the condition
\begin{equation}
 \< V_{ij ,k l } \>= 0,\quad
  \< V_{ i j, kl } V_{i'j',k'l'} \>=
  (\delta_{j,l}\delta_{j',l'}) (\delta_{i,k'} \delta_{i',k})
\label{eq:normpertur}
\end{equation}
where the $i$s and $k$s run over the environment and the coupled qubit whereas
the $i$s and $j$s run over the spectator qubit. The first set of $\delta$s
describe the fact that  the spectator is not coupled, and the second set of
$\delta$s is the usual GUE relation.  The normalization condition is no
restriction as an arbitrary factor can be absorbed in the coupling constant
$\lambda$.  However this normalization guarantees that decoherence decay is
roughly independent on the environment size, for big environments.  One could
also assume that  $V$ is well described by the GOE, however this leads to
complications (due to the weaker invariance properties of the ensemble) on
which we do not want to dwell \cite{goegue}.  We first average over the
coupling. To do so we chose a basis that diagonalizes $H_0$. In this basis,
$[\tilde V_\tau]_{i j, i' j'}= \delta_{j,j'} \rme^{\rmi t E_i} V_{i,i'} \rme^{-\rmi t
E_{i'}}$. The $i$s and $j$s run over $\mcH_\rme \otimes \mcH_\rmq$
and $\mcH_\rms$ respectively, and both $E_i$ and $E_{i'}$ are
eigenenergies of $H_0$.  As long as $V$ and $H_0$ are independent, relations
(\ref{eq:normpertur}) remain unchanged, so we can write
\begin{equation}\label{eq:VVone}
\< [ \tilde V_\tau \tilde V_{\tau'}]_{i j,i' j'} \> =
 \sum_{i''=1}^{2 N_\rme} \sum_{j''=1}^{2}
   \rme^{\rmi \tau (E_i-E_{i''}) + \rmi \tau' (E_{i''} -E_{i'})}
    \< V_{i j,i'' j''} V_{i'' j'' ,i'j'} \>=\delta_{i j,i' j'} \sum_{i''}
     \rme^{\rmi (\tau-\tau') (E_i-E_{i'})} \\
\end{equation}
and
\begin{multline}\label{eq:VVtwo}
\< [ \tilde V_\tau ]_{ij,kl} [ \tilde V_{\tau'} ]_{i'j',k'l'} \>=
 \< [ \tilde V_\tau ]_{ij,kl} [ \tilde V_{\tau'} ]_{k'l',i'j'}^* \>=\\
  \rme^{\rmi \tau (E_i-E_k) + \rmi \tau' (E_{i'}-E_{k'})}
       \<V_{ij,kl} V_{i'j',k'l'} \>
  =\delta_{ij,k'l} \delta_{kj',i'l'} \rme^{\rmi (\tau-\tau') (E_i-E_k)}.
\end{multline}
This greatly simplifies eqs.~(\ref{eq:lasas}).

\begin{figure}
\includegraphics{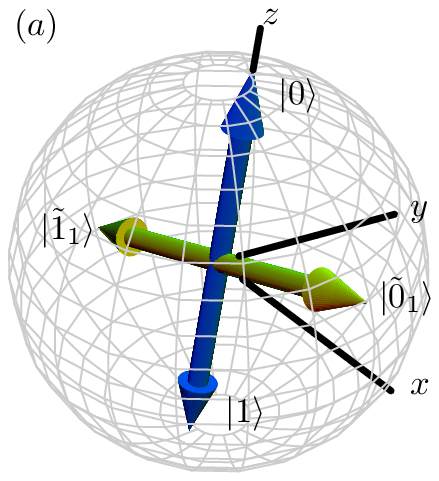}\hspace{2cm}\includegraphics{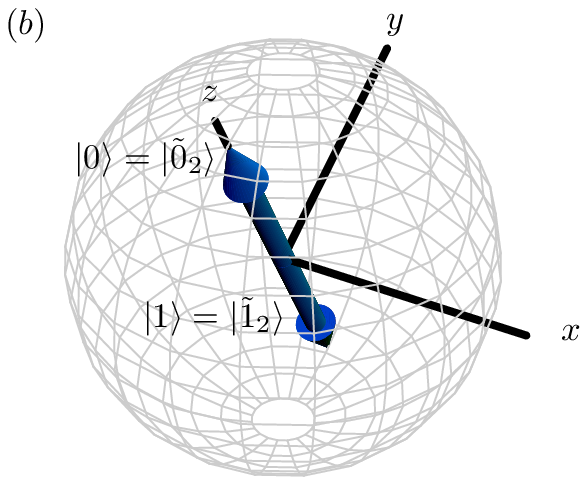}
\caption{(Color online) We present a plot to help visualize the way the
  initial condition is parametrized. On the left, the coupled qubit. The
  eigenvectors of its internal Hamiltonian ($|0\>$ and $|1\>$) are represented
  in blue. The $z$ axis is {\textit chosen} parallel to the vector $|0\>$. The
  $x$ axis is chosen in such a way as to make both $|\tilde 0_1\>$ and
  $|\tilde 1_1\>$ have real coefficients \ie such that the $xz$ plane contains
  $|\tilde 0_1\>$ and $|\tilde 1_1\>$. On the right we represent the second
  qubit where we have absolute freedom to choose the basis (even if an
  internal Hamiltonian is present), and thus we choose it according to the
  natural Schmidt decomposition.}
\label{fig:initGUEdos} \end{figure}

A further step towards obtaining the final expression is to average over the
state of the environment.  In order to ease the following work, we rewrite the
separability condition as $x_{ijk}=\psi_i \phi_{jk}$ where $|\psi_\rmc\>=\sum_i
\psi_i |i\>$ is the initial state of the environment and
$|\psi_\rmc\>=\sum_{j,k} \phi_{jk} |jk\>$ the state of the two qubits.
However, we wish to write $|\psi_\rmc\>$ in its simplest form.  We use the
invariance properties of the ensemble defined by $H_\lambda$, but under the
restriction that $H_0$ is still diagonal [to be able to write
eqs.~(\ref{eq:VVone}) and (\ref{eq:VVtwo})]. This freedom allows any unitary
operation on the purifying qubit, and a phase transformation $\exp(\rmi \alpha
H_\rmq)$ of the eigenvectors of $H_\rmq$ ($\alpha\in \Real $).  Taking into
account this freedom we now show how to construct a basis $\{|0\>,|1\>\}\otimes
\{|0\>,|1\>\}$ in which {\it an arbitrary} initial state of the central system
can be written as
\begin{equation} \label{eq:initialguewitness}
 |\psi_\rmc\>=\cos\theta_1(\cos\theta_2 |0\>+\sin\theta_2|1\>)|0\>+
 \sin\theta_1(\sin\theta_2|0\>-\cos\theta_2|1\>)|1\>,
\end{equation}
and still, $H_\rmq=\frac{\Delta}{2}|0\>\<0|-\frac{\Delta}{2}|1\>\<1|$ is
diagonal.
To find this basis we start using the Schmidt decomposition to write
\begin{equation}\label{eq:schmidt}
  |\psi_\rmc\>=\cos\theta_1|\tilde 0_1 \tilde 0_2\> +  \sin\theta_1|\tilde 1_1
      \tilde 1_2\>
\end{equation}
with $\{ |\tilde 0_i \>, |\tilde 1_i \> \}$ being an orthonormal basis of
particle $i$. For the coupled qubit, we fix the $z$ axis of the Bloch sphere
(containing both $|0\>$ and $|1\>$) parallel to the eigenvectors of $H_\rmq$,
and the $y$ axis perpendicular (in the Bloch sphere) to both the $z$ axis and
$|\tilde 0_1 \>$. The states contained in the $xz$ plane are then real
superpositions of $|0\>$ and $|1\>$, which implies that $|\tilde 0_1\> =
\cos\theta_2|0\> + \sin\theta_2|1\>$ and $|\tilde 1_1\> = \sin\theta_2|0\> -
\cos\theta_2|1\>$ for some $\theta_2$.  In the second qubit it is enough to set
$|0\>=|\tilde 0_2 \>$ and $|1\>=|\tilde 1_2 \>$. This freedom is also related
to the fact that purity only depends on $\rho_\rmq= \tr_\rms |\psi_\rmc\>
\<\psi_\rmc|$. A visualization of this procedure is found in
\fref{fig:initGUEdos}. The angle $\theta_1 \in [0,\pi/4]$ measures the
entanglement between the coupled qubit and the spectator space whereas the
angle $\theta_2 \in [0,\pi/2]$ is related to an initial magnetization of the
coupled qubit.

We have then that
\begin{align}
\rmRe \< A_J \> &= \rmRe \sum_{i,i'=1}^{N_\rme}\sum_{j,j',k,k'=0}^1
                  |\psi_i|^2 |\psi_{i'}|^2 |\phi_{jk}|^2 |\phi_{j'k'}|^2
                  \sum_{i''=1,j''=0}^{N_\rme,1}
                  \rme^{\rmi (\tau-\tau')(E_{ij}-E_{i''j''})}\nonumber \\
        & \approx
          \rmRe \sum_i \left( |\psi_i|^2 \sum_{i''}
             \rme^{\rmi (\tau-\tau')(E_{i}-E_{i''})} \right)
          \rmRe \sum_{jk} \left( |\phi_{jk}|^2  \sum_{j''}
             \rme^{\rmi (\tau-\tau')(E_{j}-E_{j''})}\right)\nonumber \\
        &\approx
           \frac{1}{N_\rme}\left( \rmRe \sum_{i,i''}
                   \rme^{\rmi (\tau-\tau')(E_{i}-E_{i''})}\right)
           \left( \sum_{jk} |\phi_{jk}|^2\rmRe
             \rme^{\rmi (\tau-\tau')E_{j}}\right)
             \left( \sum_{j''}\rme^{\rmi (\tau-\tau')E_{j''}} \right)
             \nonumber \\
        &=2 \cos^2 \left[ \frac{(\tau-\tau')\Delta}{2} \right]  \rmRe K_2(\tau-\tau')\\
        &= \rmRe K_2(\tau-\tau') \left[ 1+\cos(\Delta (\tau-\tau') \right]
\end{align}
Several approximations have been done. In the first step we assume that
the imaginary contribution from the environment is
very small. This is justified noticing that the form factor
of a random matrix ensemble is real [\eref{eq:formfactordef}]. We also take into account the normalization
condition of the states. In the second step we consider that all eigenstates
contribute similarly and replace each $|\psi_i|^2$ by its average value $1/N_\rme$.
We end up with an simple expression in the qubit (independent of the initial
condition) and the form factor [\eref{eq:formfactordef}] of the environment.
We are also assuming that the level separation in $H_\rmq$ is $\Delta$ so
$E_0=-E_1=\Delta/2$.

Studying the average over initial conditions in the environment leads to the
observation that $\< A_1 \>=\< A_2 \>=\Or(1/N_\rme)$.  The remaining term can be
computed along the same lines (\ie same techniques and assumptions),
however its calculation is both cumbersome and boring. We leave it to the
enthusiastic student. Its result is
\begin{equation}
\rmRe \< A_3 \> =\rmRe K_2(\tau-\tau')
  \left\{ 1-g_1(\theta_1,\theta_2) +
       \left[ 1-g_2(\theta_1,\theta_2) \right] \cos[\Delta (\tau-\tau')] \right \}
\label{eq:asdf}
\end{equation}
where $g_1(\theta_1,\theta_2)$ and $g_2(\theta_1,\theta_2)$ are geometric
factors that depend on the initial state. Their values are given by
\begin{align}
g_1({\theta_1,\theta_2})&= g(\theta_1) [1-g(\theta_2)]
                             + g(\theta_2) [1-g(\theta_1)]\\
g_2({\theta_1,\theta_2})&= 2[1-g(\theta_1)] - g(\theta_2)[ 1-2g(\theta_1)],
\end{align}
where $g(\theta)=\cos^4\theta+\sin^4\theta$.

The general solution for purity using this parametrization is
\begin{multline}\label{eq:generalGUEtwo}
 P(t)= 1 -2\lambda^2\int_0^t\rmd\tau\int_0^\tau\rmd\tau'
[ 1+\delta (\tau'/\tau_\rmH)-b_2 (\tau'/\tau_\rmH)] \\
\times
  \big [ g_1(\theta_1,\theta_2) +
 g_2(\theta_1,\theta_2) \cos\Delta\tau' \big ] +
 \Or(\lambda^4) .
\end{multline}

Two meaningful limits can be studied, the degenerate limit, in which the
spacing of the internal Hamiltonian is much smaller than the inverse of
the Heisenberg time of the environment ($\Delta \ll 1/\tau_\rmH$),
and the fast limit ($\Delta\gg 1/\tau_\rmH$) in which the opposite
condition is met.
In the degenerate limit purity decay is given by
\begin{equation}\label{eq:spectatorDegenerate}
 P_{\rm D}(t)=1-\lambda^2(2-g_{\theta_1})r(t),
\end{equation}
where
\begin{equation}
r(t) = t \max \{t,\tau_\rmH \} + \frac{2}{3\tau_\rmH} \min \{ t,\tau_\rmH\}^3.
\label{eq:algoasdf}
\end{equation}
The result is independent of $\theta_2$ since a degenerate Hamiltonian
is, in this context, equivalent to no Hamiltonian at all.
The $\theta_1$-dependence in this formula shows that an entangled qubit
pair is
more susceptible to decoherence than a separable one.
In the fast limit we obtain
\begin{equation}\label{eq:spectatorFast}
P_{\rm F}(t)=1-\lambda^2 \left[ g_1(\theta_1,\theta_2)r(t)
 + 2\tau_\rmH g_2(\theta_1,\theta_2) t \right] .
\end{equation}
For initial states chosen as eigenstates of $H_\rmq$ we find linear decay of
purity both below and above Heisenberg time. In order for $\rho_\rmq$ to be an
eigenstate of $H_\rmq$ it must, first of all, be a pure state (in $\mcH_\rmq$).
Therefore this behavior can only occur if $\theta_1=0$ or $\theta_1= \pi/2$.
Apart from that particular case, we observe in both limits, the fast as well as
the degenerate limit, the characteristic linear/quadratic behavior before/after
the Heisenberg time similar to the behavior of fidelity decay.

\begin{figure}
\centering \includegraphics[width=.8\textwidth]{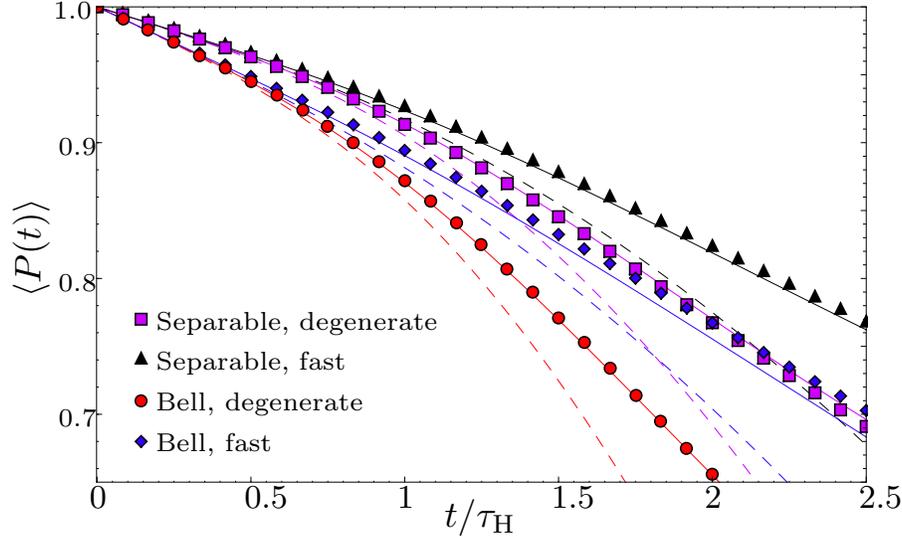}
 \caption{(Color online) Numerical simulations for the average purity as a
   function of time in units of the Heisenberg time of the environment
   (spectator configuration, GUE case).  For the coupling strength $\lambda=
   0.03$ we show the dependence of $\< P(t)\>$ on the level splitting
   $\Delta$ in $H_1$ and on the initial degree of entanglement between the
   two qubits (in all cases $\theta_2= \pi/4$): $\theta_1= 0$ (separable
   states), $\Delta=8$ (black triangles); $\theta_1= \pi/4$ (Bell states),
   $\Delta=8$ (blue rhombus); $\theta_1= 0$, $\Delta= 0$ (purple squares);
   $\theta_1= \pi/4$, $\Delta= 0$ (red circles). We also show the
   corresponding linear response results (dashed lines) and exponentiated
   linear response results (solid lines).  They are plotted with the same
   color, as the respective numerical data.  In all cases $N_\rme=1024$.}
\label{fig:holetwo}
\end{figure}

In Fig. \ref{fig:holetwo} we show numerical simulations for the average purity
decay.  We average over 30 different Hamiltonians each probed with 45 different
initial conditions. We contrast Bell states ($\theta_2=\pi/4$,
$\theta_1=\pi/4$) with separable states ($\theta_2=\pi/4$, $\theta_1= 0$), and
also systems with a large level splitting ($\Delta= 8$) in the first qubit with
systems having a degenerate Hamiltonian ($\Delta= 0$). The results presented in
this figure show that entanglement generally enhances decoherence.  This can be
anticipated since for fixed $\theta_2$, increasing the value of $\theta_1$ (and
hence entanglement) increases both $g_1(\theta_1,\theta_2)$ and
$g_2(\theta_1,\theta_2)$. At the same time we find again that increasing
$\Delta$ tends to reduce the rate of decoherence, while a strict inequality
only holds among the two limiting cases (just as in the one qubit case). From
$g_2(\theta_1,\theta_2)=2-g_1(\theta_1,\theta_2)-g(\theta_1)$, it follows that
$(P_{\rm F}-P_{\rm D})/\lambda^2=g_2(\theta_1,\theta_2) [r(t)
-2t\tau_\rmH] \ge 0$. Therefore, for fixed initial conditions and $t$ greater
than 0, $P_{\rm F}(t) > P_{\rm D}(t)$. This is the second aspect illustrated in
\fref{fig:holetwo}.

The linear response formulae derived throughout this section are expected to
work for high purities or, equivalently, when the second order Born expansion
of the echo operator is a good approximation; it is of obvious
interest to extend them to cover a larger range. The extension of fidelity
linear response formulae has, in some cases, been done with some effort using
super-symmetry techniques. This has been possible partly due to the simple
structure of the fidelity amplitude, but trying to use this approach for a more complicated
object such as purity seems to be out of reach for the time being.
Exponentiating the formulae obtained from the linear response formalism has
proven to be in good agreement with the exact super-symmetric and/or numerical
results for  fidelity, if the perturbation is not too big. The exponentiation
of the linear  response  formulae  for  purity  can  be  compared  with  Monte
Carlo simulations in order to prove its  validity.  We now explain the
details required to implement this procedure.

Given a linear response formula $P_{\rm LR}(t)$ for which $P(0)=1$
[like eqs.~(\ref{eq:generalGUEtwo}), (\ref{eq:spectatorDegenerate}), or
(\ref{eq:spectatorFast})], and an
expected asymptotic value for infinite time $P_\infty$, the exponentiation
reads
\begin{equation}
 \label{eq:ELRextension}
 P_{\rm ELR}(t)=P_\infty+(1-P_\infty)\exp\left[- \frac{1-P_{\rm
LR}(t)}{1-P_\infty}\right].
\end{equation}
This particular form guaranties that $P_{\rm ELR}$ equals $P_{\rm LR}$ for
short times, and that $\lim_{t \to \infty}P_{\rm ELR}(t)=P_\infty$ The
particular value of $P_\infty$ will depend on the physical situation; in our
case it will depend on the configuration and on the initial conditions.

Consider the spectator configuration. Let us write the initial condition as
\begin{equation}
 |\psi_{\rmc}(\theta)\>=\cos\theta_1|\tilde 0_\rmq
\tilde0_\rms\>+\sin\theta_1|\tilde 1_\rmq \tilde 1_\rms\>
\end{equation}
for some rotated basis.
We assumed that for long enough
times,
the totally depolarizing channel $\mathcal E_{\rm d}$ will act (recall
that the
totally depolarizing channel is defined as $\mathcal E_{\rm
d}[\rho]=\openone/\tr \openone$ for any density matrix $\rho$)
on the coupled qubit.
Hence, for the spectator configuration, the final value of purity is
assumed to be
\begin{equation}
P_\infty=
 P\left( \mathcal E_{\rm d}\otimes \openone
[|\psi_{\rmc}(\theta)\>\<\psi_{\rmc}(\theta)|] \right)
=\frac{g(\theta_1)}{2}.
\end{equation}
Good   agreement   is  found   with   Monte Carlo   simulations  for  moderate
and strong couplings, see fig~\ref{fig:holetwo}. For weak coupling $P_\infty$
depends on the coupling strength.

\subsection{Two qubit concurrence decay}

\begin{figure}
\centering
\psfrag{a}{}\psfrag{C}{$C$}\psfrag{P}{$P$}
\psfrag{dimHe}{$\dim \mcH_\rme$}
\psfrag{8}{8}
\psfrag{16}{16}
\psfrag{32}{32}
\psfrag{64}{64}
\psfrag{128}{128}
\psfrag{256}{256}
\includegraphics{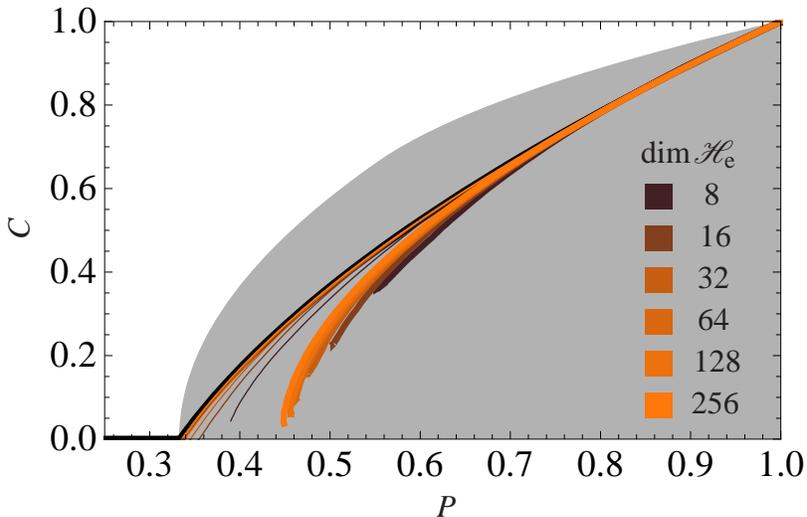}
\caption{(Color online) We present the concurrence-purity $CP$ plane. The area
 allowed for physical states is the gray area plus the set $\{(0,P),
 P\in[1/4,1/3]\}$.  The Werner curve \eref{eq:goodexponwernerCtime}  is shown
 as a thick black solid line.  We show curves $(\<C(t)\>,\<P(t)\>)$ as
 obtained from numerical simulations of the spectator Hamiltonian. We choose
 $\Delta=1$ and vary $\dim(\mcH_\e)$ for two different coupling strengths
 $\lambda=0.02$ (thick lines) and $\lambda=0.14$ (thin lines). The resulting
 curves are plotted with different colors, according to $\dim(\mcH_\e)$ as
 indicated in the figure.}
\label{fig:exampleacuuWerner_one}
\end{figure}
In the previous section we studied purity decay of $n$ qubits.  Purity
measures entanglement with the environment, but we can wonder how decoherence
affects the internal quantum mechanical properties of the central system.
Possibly the most important quantum mechanical property of a multi-particle
system is its internal entanglement.  We consider the simples scenario in
which it is possible to have entanglement: a two qubit subsystem. Two qubit
entanglement can measured with the concurrence $C$, a measure used
extensively in the literature~\cite{andrereview} that is straightforward to
calculate. It is closely related to the entanglement of formation, which
measures the minimum number of Bell pairs needed to create an ensemble of
pure states representing the state to be studied~\cite{firstconcurrence}.
Given a density matrix $\rho$ representing the state of two qubits,
concurrence is defined as
\begin{equation}
 \label{eq:concurrence}
 C(\rho)=\max \{0,\Lambda_1-\Lambda_2-\Lambda_3-\Lambda_4 \}
\end{equation}
where $\Lambda_i$ are the eigenvalues of the matrix $\sqrt{\rho (\sigma_y
\otimes \sigma_y) \rho^* (\sigma_y \otimes \sigma_y)}$ in non-increasing
order.  The superscript $^*$ denotes complex conjugation in the computational
basis and $\sigma_y$ is a Pauli matrix~\cite{wootters}. The concurrence has a
maximum value of one for Bell states, and a minimum value of zero for
separable states.  Furthermore, it is invariant under bi-local unitary
operations.  However, since concurrence is defined in terms of the
eigenvalues of a hermitian $4\times4$-matrix, an analytical treatment, even
in linear response approximation, is much more involved than in the case of
purity, and has not been performed.

We shall first explore a relation (first found in \cite{pineda:012305}, partly
explained in \cite{ziman:052325}, and further studied in \cite{pinedaRMTshort,
pinedalong}) between concurrence and purity. We show that this relation is
valid in a wide parameter range.  Combining it with the
appropriate formula for purity decay,  we obtain an analytic prediction for
concurrence decay. We compare our prediction with Monte
Carlo simulations. It is essential that the two qubits do not interact;
otherwise the coupling between the qubits would act as an additional sink (or
source) for internal entanglement -- a complication we wish to avoid.

\begin{figure}
\centering
\psfrag{q}[cb]{$\log_2 N_\rme$} \psfrag{dimHe}[lb]{$N_\rme$}
\psfrag{E}[c]{$\log_2 D$}\psfrag{a}{(a)}\psfrag{b}{(b)}
\psfrag{d}[cb]{$\lambda$}
\psfrag{l1}{$\lambda=0.14$}
\psfrag{l2}{$\lambda=0.02$}
\includegraphics[height=5.5cm]{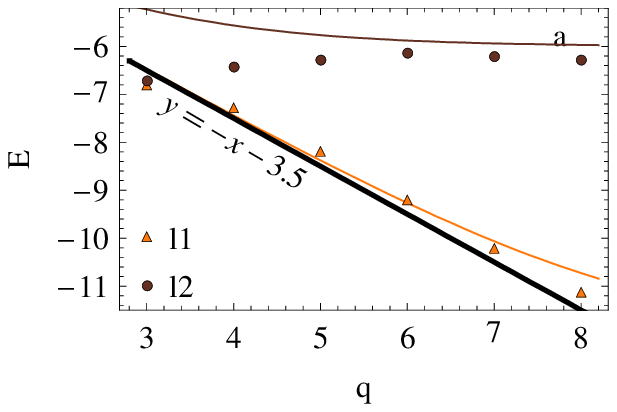}\includegraphics[height=5.5cm]{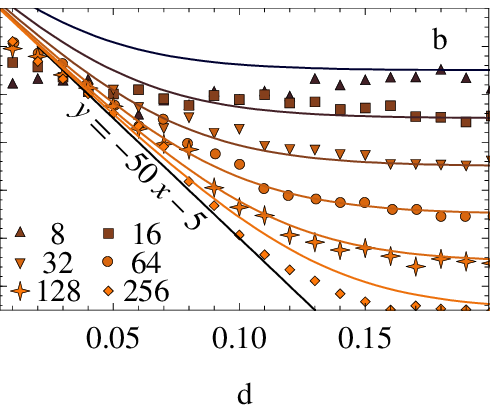}
\caption{(Color online) We show $D$ [\eref{eq:deferrorwerner}] which measures
  a ``distance'' in the $CP$ plane between numerical curves for the RMT model
  and the Werner curve. In (a) its dependence on the size of the environment
  is studied for two cases. For $\lambda=0.02$  a finite value is approached,
  whereas for $\lambda=0.14$  algebraic decay is seen. The thick line, which
  is proportional to $N_\rme^{-1}$, is meant as a guide to the eye.  In (b) we
  plot $D$ as a function of the coupling $\lambda$, for various values of
  $N_\rme$.  In the large environment limit, and for $\lambda \ll 0.1$, we
  have a noticeable deviation from the Werner curve. In all cases $\Delta=1$.}
\label{fig:transicion}
\end{figure}

We study the relation between concurrence and purity using the $CP$-plane,
where we plot concurrence against purity with time as a parameter.
This plane is plotted in \fref{fig:exampleacuuWerner_one}. The gray area
indicates the region of physically admissible states \cite{munro:030302}.  The
upper boundary of this region is given by the maximally entangled mixed states.
The Werner states $\rho_{\rm W}=\alpha \frac{\openone}{4}+ (1-\alpha)
|\rm{Bell}\>\<\rm{Bell}|$, $0\le \alpha\le 1$, define a smooth curve on the
$CP$-plane (black solid line).  The analytic form of this curve is
\cite{pinedaRMTshort}
\begin{equation}\label{eq:goodexponwernerCtime}
C_{\rm W}(P)=\max\left\{0, \frac{\sqrt{12P-3}-1}{2}\right\},
\end{equation}
and will be referred to as the Werner curve. However, note that states mapped
to the Werner curve are not necessarily Werner states.  We perform numerical
simulations in the spectator configuration.  We compute the average
concurrence for a given interval of purity using 15 different Hamiltonians
and 15 different initial states for each Hamiltonian. We fix the level
splitting in the coupled qubit to $\Delta=1$ and consider two different
values $\lambda=0.02$ and $0.14$ for the coupling to the environment.
\Fref{fig:exampleacuuWerner_one} shows the resulting $CP$-curves for
different dimensions of $\mcH_\e$.  Observe that for both values of
$\lambda$, the curves converge to a certain limiting curve as $\dim(\mcH_\e)$
tends to infinity. While for $\lambda=0.02$, this curve is at a finite
distance of the Werner curve, for $\lambda=0.14$ it practically coincides
with $C_{\rm W}(P)$.  Varying the configuration, the coupling strength, the
level splitting, or the ensemble (GOE/GUE), gives the same qualitative
results in the $CP$ plane, for large dimensions. In some cases we have an
accumulation towards the Werner curve, in others there is a small variation.

To study this situation in more detail, consider a $CP$-curve $C_{\rm
num}(P)$, obtained from our numerical simulations, and define its
``distance'' $D$ to the Werner curve as
\begin{equation}
 \label{eq:deferrorwerner}
 D=\int_{P_{\rm min}}^1\rmd P \left| C_{\rm num}(P)-C_{\rm W}(P)\right|.
\end{equation}
The behavior of $D$ as a function of the size of the system is shown in Fig.
\ref{fig:transicion}(a). For $\lambda=0.14$ (dark dots), the error goes to
zero in an algebraic fashion, at least in the range studied.  In fact, from a
comparison with the black solid line we may conclude that the deviation $D$
is inversely proportional to the dimension of $\mcH_\e$.  By contrast, for
$\lambda= 0.02$ (light red dots), $D$ tends to a finite value, in line with
the assertion that the numerical results converge to a different curve.  In
Fig. \ref{fig:transicion}(b) we plot the error $D$ as a function of
$\lambda$, for different dimensions of $\mcH_\e$.  The results suggest an
exponential decay of $D$ with the coupling strength.  The simplest dependence
of the deviation in agreement with these two observations is
\begin{equation}\label{eq:depE}
 D_{\Delta=1}=\frac{1}{2^{3.5}N_\rme}+\frac{1}{2^{5+50 \lambda}}.
\end{equation}
We also plot the curves corresponding to this ansatz in
\fref{fig:transicion}. Good agreement is observed for $D \ll 1$.  Notice the
exponential decay of $D$ with respect to $\lambda$. One can thus, in an
excellent approximation for large $\lambda$, say that for large dimensions
the limiting curve is the Werner curve.  For $\Delta=0$, all studied couplings
numerical convergence to the Werner curve was observed in the large $N_\rme$
limit.

\begin{figure}
\centering
\psfrag{a}{}
\psfrag{b}{(b)}
\psfrag{C}{$C$}
\psfrag{P}{$P$}\psfrag{toth}[b]{$t/\tau_\rmH$}
\includegraphics{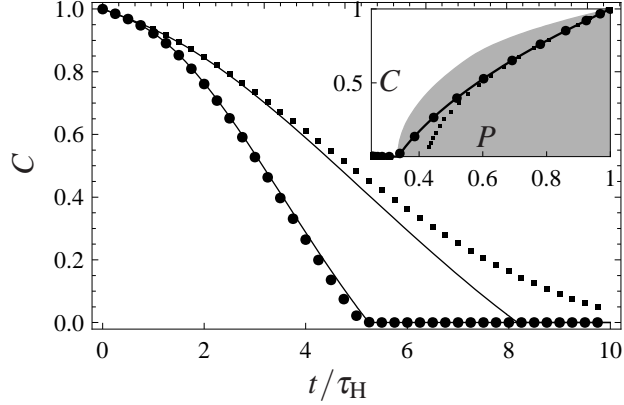}
\caption{We show numerical simulations for the average concurrence as a
  function of time when both qubits are coupled with equal strength $\lambda$
  to the environment.  We consider small couplings $\lambda=0.01$ which lead
  to the Gaussian regime for purity decay. The large dots show the result
  without internal dynamics, whereas the small ones are obtained for fast
  internal Rabi oscillations ($\Delta=10$ for both qubits). The theoretical
  expectation for concurrence decay based on \eref{eq:concurrenceELR} are
  also plotted.  The inset displays the corresponding
  evolution in the $CP$-plane with the same symbols as used in the main
  graph. In all cases $N_\e=1024$}
\label{fig:CPdecay}
\end{figure}

Sufficiently close to $P=1$, the above arguments imply a one to one
correspondence between purity and concurrence, which simply reads $C\approx
P$.  This allows to write an approximate expression for the behavior of
concurrence as a function of time
\begin{equation}
 \label{eq:concurrenceLR}
 C_{\rm lr}(t)=P_{\rm LR}(t),
\end{equation}
using the appropriate linear response result for purity decay. The
corresponding expressions for purity decay have been discussed in detail
above. \Eref{eq:concurrenceLR} has similar limits of validity
as the linear response result for the purity.  As it implies this
approximation, we call it the linear response expression
for concurrence decay.

In those cases where the deviation from the Werner curve [see
\eref{eq:goodexponwernerCtime}] is small and where the exponentiated linear
response expression holds for the average purity,
we can write down a phenomenological formula for concurrence decay, which is
valid over the whole range of the decay
\begin{equation}
 \label{eq:concurrenceELR}
 C_{\rm elr}(t)=C_{\rm W}(P_{\rm ELR}(t)) \; .
\end{equation}
In Fig. \ref{fig:CPdecay} we show random matrix simulations
when both qubits are coupled with equal strength $\lambda$ to an environment.
We consider small couplings
$\lambda_1=\lambda_2= 0.01$ which lead to the Gaussian regime for purity
decay. We find good agreement with the prediction of
\eref{eq:concurrenceELR}, except when we switch-on a
fast internal dynamics in both qubits ($\Delta= 10$) and
consequently $D$ is large. See the inset in \fref{fig:CPdecay}.

A similar analysis can be carried if one  starts  with a non-Bell state.
However, as the reader can suspect this introduces more difficulties and
more complicated expressions. The details can be found in \cite{pinedalong}.

\section{Conclusions and outlook}

The stability of quantum computation is the core of this course. We have given
models for both fidelity decay and decoherence, based on random matrix theory.
The most common exponential decay is a standard limit in both cases, but we
have shown, that RMT not only reproduces the usual results in these cases, but
provides a richer spectrum of behavior. Basically it introduces the Heisenberg
time (of the total system or of the environment) as an additional time scale.
This time scale in the case of fidelity decay defines the transition between
quadratic (Gaussian)  or perturbative and linear (exponential) or Fermi
golden rule decay. Both regimes were known but with RMT the transition is
well described. Furthermore it marks a clear difference to the behavior of
integrable systems, where the size of the effective Hilbert space tends to be
much smaller. This leads to faster decay, as the onset of quadratic behavior
is much earlier.

This brings us to an interesting discussion of great practical
importance. Chaotic perturbations seem to be less harmful than integrable
ones.  This statement has given rise to many discussions with
experimentalists \cite{Pastawskiprivate2007}, who, with good reason, say that
the argument is doubtful because, with some adequate pulse in quantum optics
or NMR (nuclear magnetic resonance), an integrable perturbation can be
corrected for.  Discussions of such corrections are found, e.g., in
\cite{Berman01, Gorin-Lopez}.

Yet there is an important alternate aspect: By chaotizing an algorithm (after
doing all the corrections possible) one can reach additional stabilization.
This was first discussed for a quantum Fourier transform by Prosen and {\v
Z}nidari{\v c} \cite{0305-4470-35-6-309}. They showed, that by introducing
additional gates they can decrease correlations between perturbations and by
consequence slow down fidelity decay.  The idea to use random matrix theory
to optimize such a method \cite{ShepelyanskyRMTFidelity} took hold later and
opens up very attractive alternatives. The basis of the idea is to randomly
change the computational basis between gates. This field is just developing,
and we refer the reader to \cite{kern:062302} for recent advances. Fidelity
freeze provides an additional option to improve fidelity decay. If we can
show that pulse manipulation can eliminate, or reduce, the diagonal part of
the perturbation, a freeze situation can be induced. The fact that even in
the absence of strong spectral correlations a freeze can be typical, if
not the ensemble average, opens a new alley of research.

\begin{theacknowledgments} We thank Thomas Gorin and Toma\v z Prosen for for
many useful discussions. We acknowledge support from
the grants PAPIIT IN112507 and CONACyT 57334.  
\end{theacknowledgments}

\bibliographystyle{aipproc}
\bibliography{paperdef,miblibliografia,lorev}

\end{document}